\documentclass[aip,jcp,reprint,showpacs,amsmath,amssymb,floatfix]{revtex4-1}

\usepackage{tikz}
\usetikzlibrary{calc}
\usepackage{amsmath}
\usepackage{graphicx}
\usepackage{hyperref}

\usepackage{upgreek}
\begin{document}

\renewcommand*{\theHtable}{\arabic{table}} 
\renewcommand*{\theHfigure}{\arabic{figure}} 

\title{Introducing Improved Structural Properties and Salt Dependence into a Coarse-Grained Model of DNA}

\author{Benedict E.~K.~Snodin}
\email{benedict.snodin@chem.ox.ac.uk}
\affiliation{Physical and Theoretical Chemistry Laboratory, Department of Chemistry, University of Oxford, South Parks Road, Oxford, OX1 3QZ, United Kingdom}
\author{Ferdinando Randisi}
\affiliation{Life Sciences Interface Doctoral Training Center, South Parks Road, Oxford, OX1 3QU, United Kingdom}
\affiliation{Rudolf Peierls Centre for Theoretical Physics, 1 Keble Road, Oxford OX1 3NP, United Kingdom}
\author{Majid Mosayebi}
\affiliation{Physical and Theoretical Chemistry Laboratory, Department of Chemistry, University of Oxford, South Parks Road, Oxford, OX1 3QZ, United Kingdom}
\author{Petr \v{S}ulc}
\affiliation{Center for Studies in Physics and Biology, The Rockefeller University, 1230 York Avenue, New York, NY 10065, USA}
\author{John S.~Schreck}
\affiliation{Physical and Theoretical Chemistry Laboratory, Department of Chemistry, University of Oxford, South Parks Road, Oxford, OX1 3QZ, United Kingdom}
\author{Flavio Romano}
\affiliation{Physical and Theoretical Chemistry Laboratory, Department of Chemistry, University of Oxford, South Parks Road, Oxford, OX1 3QZ, United Kingdom}
\author{Thomas E.~Ouldridge}
\affiliation{Department of Mathematics, Imperial College, 180 Queen's Gate, London SW7 2AZ, United Kingdom}
\author{Roman Tsukanov}
\affiliation{Department of Chemistry and the Ilse Katz Institute for Nanoscale Science and Technology, Ben-Gurion University of the Negev, Beer Sheva, Israel}
\author{Eyal Nir}
\affiliation{Department of Chemistry and the Ilse Katz Institute for Nanoscale Science and Technology, Ben-Gurion University of the Negev, Beer Sheva, Israel}
\author{Ard A.~Louis}
\affiliation{Rudolf Peierls Centre for Theoretical Physics, 1 Keble Road, Oxford OX1 3NP, United Kingdom}
\author{Jonathan P.~K.~Doye}
\email{jonathan.doye@chem.ox.ac.uk}
\affiliation{Physical and Theoretical Chemistry Laboratory, Department of Chemistry, University of Oxford, South Parks Road, Oxford, OX1 3QZ, United Kingdom}

\onecolumngrid

\begin{abstract}
We introduce an extended version of oxDNA, a coarse-grained model of DNA designed to capture the thermodynamic, structural and mechanical properties of single- and double-stranded DNA. By including explicit major and minor grooves, and by slightly modifying the coaxial stacking and backbone-backbone interactions, we improve the ability of the model to treat large (kilobase-pair) structures such as DNA origami which are sensitive to these geometric features. Further, we extend the model, which was previously parameterised to just one salt concentration ($[\text{Na}^+]=0.5\,\text{M}$), so that it can be used for a range of salt concentrations including those corresponding to physiological conditions. Finally, we use new experimental data to parameterise the oxDNA potential so that consecutive adenine bases stack with a different strength to consecutive thymine bases, a feature which allows a more accurate treatment of systems where the flexibility of single-stranded regions is important. We illustrate the new possibilities opened up by the updated model, oxDNA2, by presenting results from simulations of the structure of large DNA objects and by using the model to investigate some salt-dependent properties of DNA.
\end{abstract}

\maketitle

\twocolumngrid
\section{Introduction}
Deoxyribonucleic acid (DNA) performs the crucial function of storing genetic information in living organisms. It is made up of repeating units called nucleotides, each of which consists of a sugar and phosphate backbone plus a base (either Adenine (A), Guanine (G), Thymine (T), or Cytosine (C)) attached to the sugar. Watson-Crick base-pairing, A with T and G with C, along with planar stacking interactions between bases and the constraints of the backbone, leads to the formation of the well-known double-helical structure of DNA. 

The Watson-Crick complementarity of DNA also permits the rational design of DNA objects for which the intended structure is the global free-energy minimum, a property which has been exploited to create a wide variety of 2D and 3D nanostructures.\cite{3dDietz2009,yanGridiron2013,pengYinLego2012} Further, these DNA objects can be functionalised,\cite{tetrahedronToCell2011,douglasNanorob2012,liedlGoldNanoOrigami2012,dnaBox2012} with potential applications ranging from nanomedicine to nanoelectronics.\cite{yanReview2011}

Theoretical and computational approaches to modelling DNA have been widely exploited to probe the behaviour of the molecule in both a biological and a nanotechnological context. At the finest level of detail, quantum chemistry calculations have been used to study the interactions between nucleotides,\cite{Svozil10,sponerQChem2013,sponerQChem2013prac} although the high computational cost of such an approach limits these methods to interactions between nearest-neighbour base pairs in vacuum. Classical all-atom approaches, where every atom of DNA and the surrounding solvent is modelled as a point particle with effective interactions, have been widely employed to study small DNA motifs,\cite{atomisticRev2011,Perez2012} and have recently been applied to larger DNA systems.\cite{aksimentiev2012,Yoo2013,aksimentievAllAtomReview2014} However, simulating rare-event processes such as the breaking or formation of base pairs remains a challenge with these models, with $\upmu s$ time scales being the limit of what is currently accessible.\cite{aksimentievAllAtomReview2014} At the other end of the scale, theoretical approaches have been developed to understand certain large-scale properties of DNA. These include the wormlike-chain model, which treats DNA as a continuously flexible polymer.\cite{wormlikeChain1994} While such models can provide useful insights into the physical properties of DNA, they are not detailed enough, by design, to address processes such as duplex formation.

The middle ground between analytical and all-atom approaches is occupied by coarse-grained DNA models. Such models integrate out many of the degrees of freedom of the DNA nucleotide, and often neglect the solvent molecules; these approximations inevitably imply a compromise between accuracy, generality and computational efficiency, so that care must be taken in applying these models to a given problem. However, the simplified picture presented by such an approach can be a strength, as, in addition to greatly increasing the time scale and number of nucleotides that can be studied, it can allow one to understand the generic physics governing the system more easily.

Many coarse-grained DNA models have been developed in recent years;\cite{hinckley2013experimentally,Papoian10,Cragnolini13,meanFieldDuplex,Linak11,Morriss-Andrews2010,Araque11,lyubartsev2014,maffeoCGssDNA2014} in this work we improve the model due to Ouldridge \textit{et al.}, oxDNA.\cite{dnaModelStruct,ouldridgeThesis} This nucleotide-level model is designed with a heuristic, ``top-down'' approach, with a focus on reproducing well-known properties of DNA (such as the helical structure of the B-DNA duplex) and experimental results (such as duplex melting temperatures), rather than, for example, building the model up by integrating out details from an all-atom representation. In the oxDNA model, each nucleotide is a rigid body with three interaction sites that have mutual, highly anisotropic interactions. This treatment is sufficient to obtain good agreement with experimental data on the structural, mechanical and especially the thermodynamic properties of single- and double-stranded DNA. Consequently, the model has provided key insights into many different processes relevant to DNA nanotechnology\cite{Ouldridge_tweezers_2010,necitujsulc2012simulating,ouldridge2013dna,ouldridge2013optimizing,dnagels,schreckBulge,schreckHairpin,srinivas2013biophysics,majidRupture2015} and biophysics,\cite{Matek,wang2014modeling,matekPlectoneme,majidLoopStacking,romano2013coarse} and importantly, has also been shown to provide direct agreement with experimentally measured properties on a range of systems including DNA overstretching,\cite{romano2013coarse} a two-footed DNA walker,\cite{ouldridge2013optimizing} and toehold-mediated strand displacement.\cite{srinivas2013biophysics,machinekDisplacement2014}

Despite these achievements, there are some areas where oxDNA can be improved. The model was parameterised to $[\text{Na}^+]=0.5\,\text{M}$, a high salt concentration similar to that used for many applications in DNA nanotechnology. However, the ability to study DNA behaviour as a function of salt would allow quantitative comparison with a greatly expanded set of experiments, and, in particular, if we wish to apply oxDNA to biological systems we would like to work at physiological salt ($[\text{Na}^+]\approx 0.15\,\text{M}$). At the same time, the wealth of experimental data for the thermodynamic and mechanical properties of DNA as a function of salt concentration\cite{ritort2013ssDNAsalt,deniz2014saltHairpin,SantaLucia2004,taylor1990cyclisation} makes fitting the salt-dependent properties of an extended model possible.

While the detailed effect of salt on DNA electrostatics can be highly complex,\cite{mocciSalt2012,Potoyan2013} here we use a simple Debye-H\"uckel interaction term, as first implemented for oxDNA by Wang and Pettitt in Ref.~\onlinecite{wang2014modeling}, to model how salt screens repulsive interactions. The more coarse-grained description is commensurate with the level of approximation used more generally in the oxDNA framework. We carefully parameterise the model to an extended set of experimental data for melting temperatures and persistence lengths, including the behaviour of single strands.

A second area that merits attention is the performance of the model in simulating the structure of large (kilobase-pair) DNA objects. The model value of the B-DNA pitch did not come under too much scrutiny in the original parameterisation as there is some disagreement in the literature about the precise value of the average pitch.\cite{taylor1990cyclisation,shore1983,dawidDNAPitch,ido2013pitch,du2005cyclisation,wang1979helicalRepeat} However, improved experimental techniques in fabrication and imaging of large DNA objects\cite{tetrahedronCryoEM2009,3dDietz2009,dietzCryoEM2012} have presented an opportunity to finely tune this value, because small adjustments to the duplex pitch can result in significant changes to the global twist of a large-scale DNA nanostructure. In addition, simulating these large-scale structures has illustrated the potential importance of nicks and junctions for the effective duplex pitch, so that improving the model's description of these effects has become a priority. Finally, the original model duplex had grooves with equal widths, whereas B-DNA is known to have a larger major groove and a smaller minor groove. This implies that the positions of the backbone sites in the model, which are directly related to the groove widths, could be more realistic. This detail could be relevant, for example, in origami structure, where the precise backbone positions are known to be important for junction placement.\cite{3dDietz2009}

The oxDNA model was previously given sequence-dependent hydrogen-bonding and stacking strengths\cite{petrSeqDep2012} by fitting to the duplex thermodynamics of the SantaLucia model.\cite{SantaLucia2004} As the SantaLucia model gives results at the base-pair step level, one can only extract the average stacking strength for the two stacking interactions present in a given base-pair step. In particular, this means that in oxDNA the AA and TT stacking interactions are the same, whereas it is well known that the AA stacking interaction is significantly stronger than the TT interaction, an important property for DNA nanotechnology, for example, where poly-T single-stranded regions are often used as flexible linkers.\cite{Anderson08,maoztile} To remedy this we use new experimental data to reparameterise the AA and TT stacking interactions in the model.

In the following sections, after briefly introducing the original oxDNA model, we consider each change to the model in turn, and then we highlight some important aspects of the behaviour of the new model.

\section{The original oxDNA model}

In the oxDNA model introduced by Ouldridge \textit{et al.}\cite{dnaModelStruct} strands of DNA are represented by a chain of rigid bodies, with each rigid body representing a nucleotide. The coarse-grained, pairwise potential for the model can be written as a sum of interaction terms

\begin{align}
\begin{split}
V &= \sum_{\text{nearest neighbours}}(V_{\text{backbone}} + V_{\text{stack}} + V'_{\text{exc}}) \;+ \\
  & \sum_{\text{other pairs}}(V_{\text{HB}} + V_{\text{cross stack}} + V_{\text{exc}} + V_{\text{coax stack}}) \text{.} 
\end{split}
\label{eq:canonical_potential}
\end{align}

The nearest neighbours (adjacent nucleotides on a DNA strand) interact via $V_{\text{backbone}}$, $V_{\text{stack}}$, and $V'_{\text{exc}}$, which represent the connectivity between neighbouring backbones, the favourable stacking interactions between neighbouring bases, and excluded volume terms, respectively. All other nucleotide pairs interact with $V_{\text{HB}}$, $V_{\text{cross stack}}$, $V_{\text{exc}}$ and $V_{\text{coax stack}}$, corresponding to hydrogen bonding between complementary bases, cross-stacking, excluded volume, and coaxial stacking between non-nearest neighbours, respectively. We now highlight aspects of the original model that are relevant to the improvements made in this paper.

Firstly, in its original formulation, the model was parameterised for a sodium ion concentration of $0.5\,\text{M}$, chosen to reflect the high salt concentrations commonly used in DNA nanotechnology. The electrostatic interactions between the negatively charged phosphate groups on the DNA backbone were incorporated into the backbone site's excluded volume, an approximation which can be justified by the very short Debye screening length at that relatively high ion concentration. 

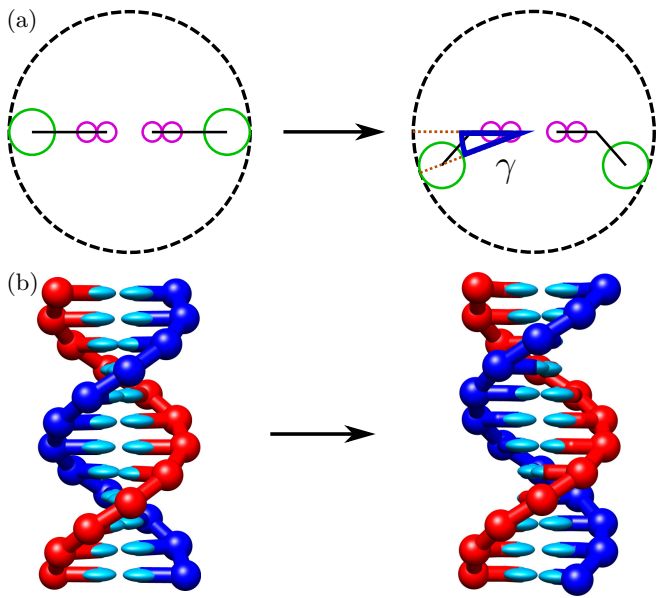
\begin{figure}
\begin{tikzpicture}
\begin{scope}
\draw node[anchor=north west] (schematic) {\includegraphics[width=\linewidth]{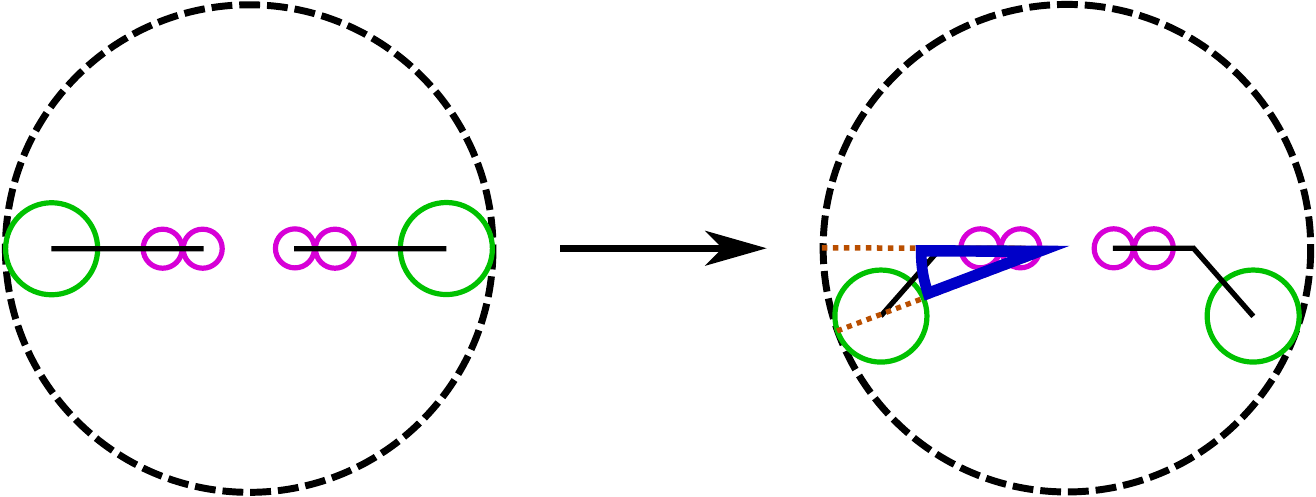}};
\draw (6.5cm,-2cm) node[anchor=north west] (gamma) {\Large$\gamma$};
\draw (schematic.north west) node[anchor=north west] (a) {(a)};
\draw ($(schematic.south west) + (0.4cm,0.0cm)$) node[anchor=north west] (oxdna) {\includegraphics[width=0.9\linewidth]{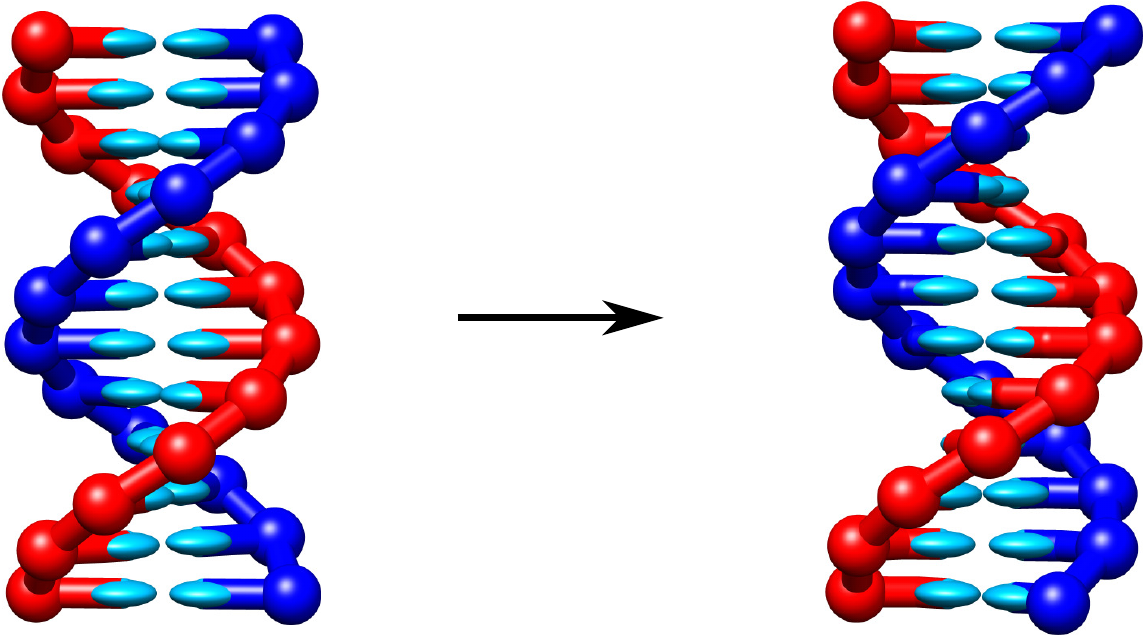}};
\draw (schematic.south west) node[anchor=north west] (b) {(b)};
\end{scope}
\end{tikzpicture}
  \caption{Schematics contrasting the original oxDNA model (left), with equal groove widths, with oxDNA2 (right), which has differentiated major and minor grooves. (a) A cross section of a duplex with one base pair displayed. The large dashed circle shows the helix radius, and each nucleotide is represented by three circles joined by a line; the large solid circles represent the backbone sites, while the small solid circles represent the stacking (closer to the backbone) and hydrogen-bonding (at the end of the nucleotide) sites. For oxDNA2, a value of 20$^{\circ}$ was chosen for the angle $\gamma$. (b) A representation of a DNA duplex for each model.}
  \label{fig:mmgroove}
\end{figure}

Secondly, the original oxDNA model\cite{dnaModelStruct} represented each nucleotide as a linear rigid body (Fig.~\ref{fig:mmgroove}). The optimal configuration for base-pairing occurs when the two nucleotides point directly at each other. As a consequence, the DNA double helix was symmetric, with the two grooves having equal widths.

Thirdly, in the original oxDNA model introduced in Ref.~\onlinecite{dnaModelStruct}, all four types of base were treated equally except that only A-T and G-C base pairs could be formed. Later \v{S}ulc \textit{et al.}\cite{petrSeqDep2012} introduced sequence-dependent thermodynamics into the model by making the strengths of the hydrogen-bonding and stacking terms depend on the identities of the interacting nucleotides. The nearest neighbour DNA model of SantaLucia,\cite{SantaLucia2004} to which oxDNA was parameterised, does not resolve the difference between AA and TT stacking as it works on the base-pair step level -- therefore AA and TT stacking strengths were set to be the same in Ref.~\onlinecite{petrSeqDep2012}.

In this work we mostly work from the original, sequence-averaged parameterisation of the model rather than the sequence-dependent one, as it is more efficient to fit the thermodynamic parameters to sequence-averaged duplex melting temperatures as given by the SantaLucia model. The exception is the parameterisation of the AA and TT stacking strengths, which did use the sequence-dependent parameters from Ref.~\onlinecite{petrSeqDep2012} as a starting point, to allow the best possible comparison between the model and the experimental results that were used for the fitting. After the parameters for oxDNA2 had been obtained, including new values for the sequence-averaged hydrogen bonding and stacking strengths, we then rescaled the sequence-dependent interaction strengths from Ref.~\onlinecite{petrSeqDep2012} accordingly for use with the new model.

\section{Introducing different widths for major and minor DNA grooves}
\label{section:major minor grooving}
B-DNA in the original oxDNA model has equal groove widths, while in reality DNA has a larger major groove and a smaller minor groove. Having realistic widths for the major and minor grooves is equivalent to having appropriately positioned backbone sites in the model, an important feature for the physical properties of many DNA motifs. For example, in DNA origami antiparallel double helices are joined by crossovers, for which the position of the backbone has been shown to be crucial for origami structure.\cite{3dDietz2009,douglas2009} Another example is anisotropic duplex bending: the duplex can be expected to bend more easily into the major groove than into the minor, if the groove widths are unequal.

The oxDNA nucleotide is composed of three interaction sites: the hydrogen-bonding, stacking and backbone sites. We introduce different groove widths by changing the position of the backbone site while keeping the duplex radius unchanged (Fig.~\ref{fig:mmgroove}), such that, rather than lying on a straight line, the three interaction sites lie in a plane. The new nucleotide shape introduces an additional parameter into the model, the angle $\gamma$ between the line from the duplex centre to the backbone site and the line from the duplex centre to the stacking site (Fig.~\ref{fig:mmgroove}(a)). Given the coarse graining of the $18$ atoms of the sugar-phosphate DNA backbone into a single interaction site, there is no definitive choice for the precise position of the backbone site and thus the value of the model parameter $\gamma$ (Fig.~\ref{fig:mmgroove}(a)). We set $\gamma=20^{\circ}$, a value which maps onto a full-atom representation of a DNA duplex well by visual inspection, although values of $\gamma$ between $15^{\circ}$ and $25^{\circ}$ would give an equally satisfying visual match.

The backbone site is moved such that the duplex radius is unchanged, and we note that the modification has a negligible added computational cost when simulating the model. However, the thermodynamic and mechanical properties are slightly affected. For the thermodynamics, we found a change of 1-2$\,$K in the duplex melting temperatures, and we modified the hydrogen-bonding and stacking strengths using the histogram reweighting method described in Section II C 2 of Ref.~\onlinecite{necitujoxRNA}, so that the agreement with experimental melting temperatures was as good as for the original model. The mechanical properties of DNA are less well constrained experimentally and so were not refitted. The mechanical properties for the new model can be found in Section \ref{section:physical properties}.

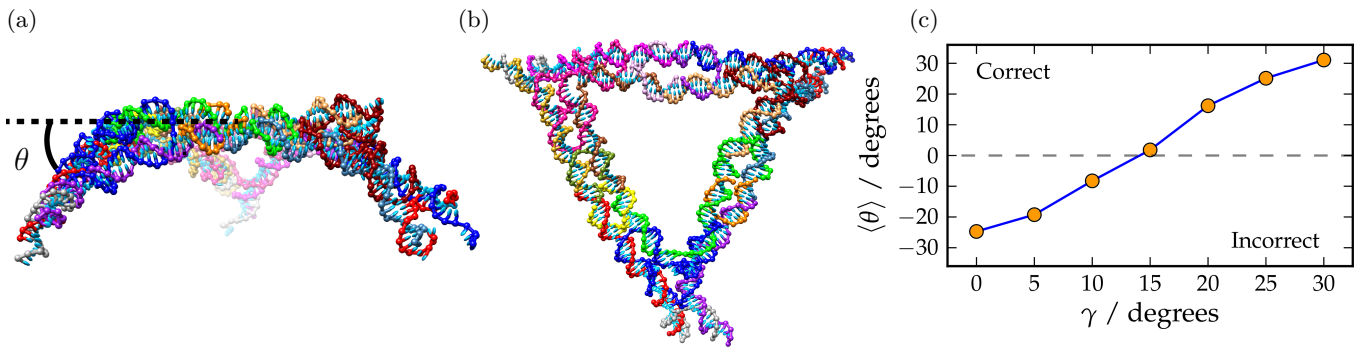
\begin{figure*}
  \begin{tikzpicture}
  \begin{scope}
  \draw node[anchor=north west] (pics) {\includegraphics[width=\linewidth]{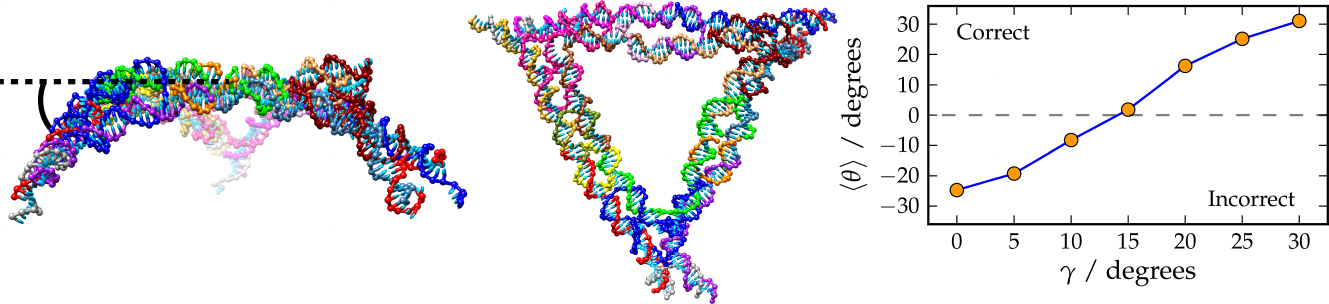}};
  \draw ($(pics.north west) + (0.0cm,0.4cm)$) node[anchor=north west] (a) {(a)};
  \draw ($(a.north west) + (0.1cm,-1.85cm)$) node[anchor=north west] (theta) {\large$\theta$};
  \draw ($(a.north west) + (6.0cm,0.0cm)$) node[anchor=north west] (b) {(b)};
  \draw ($(b.north west) + (6.0cm,0.0cm)$) node[anchor=north west] (c) {(c)};
  \end{scope}
  \end{tikzpicture}
  
  \caption{In oxDNA, the structure of the DNA trimer due to Zhang \textit{et al.}\cite{maoChiralCage2012} is found to qualitatively change as a function of duplex groove angle $\gamma$. A representative trimer configuration with $\gamma=20^{\circ}$ is shown in (a) from the side and with $\theta$, the angle between each arm and the plane of the main, triangular trimer section, displayed and in (b) from the top. (c) Shows the average value of $\theta$ as a function of $\gamma$, with $\gamma=0^{\circ}$ corresponding to equal groove widths as for the original oxDNA model, and $\gamma=20^{\circ}$ being chosen for oxDNA2.}
  \label{fig:mao trimer}
\end{figure*}

One illustration of the importance of the groove widths in oxDNA for the structural properties of DNA assemblies is provided by systems of 3-arm star tiles that are designed to form triangular prismatic polyhedra. We find that modifying the groove widths qualitatively changes the structure of trimers of these tiles (Fig.~\ref{fig:mao trimer}). Specifically, the body of the trimer defines a plane with two distinct faces. Zhang \textit{et al.}~\cite{maoChiralCage2012} found that one of two possible isomers of the polyhedron preferentially formed, implying that the free arms of the trimer systematically pointed in the direction of one of these two faces. We find a consistent result when the groove widths specified by oxDNA2 are used. When equal-sized grooves are used (as in the original oxDNA model) the trimer arms point in the opposite direction.

\section{Effective electrostatic interactions}
\label{section:dh}
One major improvement presented in this paper is the introduction of a
salt-dependent interaction term in the oxDNA model. Since
the original goal of the oxDNA model was to simulate nanotechnology experiments,
the thermodynamic and structural parameterisation was carried out at high salt.
The very short electrostatic screening length at these conditions allows one to
incorporate the electrostatics into a soft excluded volume, somewhat
circumventing the necessity of a proper treatment. The original parameterisation of
the thermodynamics was carried out at 0.5\,M [Na$^+$], a high enough value 
that further increasing it does not significantly change the physics at our
level of coarse-graining.

The problem of treating electrostatics properly for DNA in solution is a complicated
one. Perhaps the most evident issue is that the typical dimensions of
nucleotides are comparable to the Debye length of the solution, rendering a
mean-field treatment hard to justify. Also, in some cases the presence of salt
ions affects the local structure of nucleic acids, by stabilizing some
arrangements or destabilizing others. Ion condensation may also lead to 
stronger screening of the electrostatic interactions than otherwise expected -- this 
has been incorporated into coarse-grained DNA models through partial effective
charges.\cite{3spnSalt, Maffeo2010} Thus, in principle, many
non-trivial effects must be taken into account when modelling the
electrostatic interactions, and the debate on the best way to do so implicitly is still
unresolved.\cite{mocciSalt2012,Potoyan2013}

Here, we choose a very simple treatment, based on the Debye-H\"uckel model for
screened electrostatics. This approach has been used previously for other
coarse-grained models,\cite{3spnSalt,Morriss-Andrews2010} and was first introduced into oxDNA by Wang and
Pettitt.\cite{wang2014modeling} We note that this treatment is consistent with the coarse-grained nature
of the model, and we use the same top-down strategy that was used in the original
parameterisation to design the effective electrostatic interactions: our
goal is to introduce a term in the potential that will reproduce the
thermodynamic and mechanical effects of salt concentration on DNA, and
thus should be regarded as an effective interaction rather than an attempt to
rigorously model the local effects of charges. This should be kept in mind when
interpreting the results obtained with oxDNA, in particular at low salt
concentration.

Since the modelling of electrostatics is rather crude, we restrict our
parameterisation to salt concentrations of $0.1\,$M of monovalent salt or greater.
This restriction is also due to the fact that we parameterise our thermodynamics
to the model of SantaLucia,\cite{SantaLucia2004} which was fitted in a similar salt regime. 
Importantly, physiological conditions fall within this range,
which will allow quantitative comparison between simulations of DNA systems with
our model and experiments at physiological conditions.

The detailed Debye-H\"uckel form which is added to the non-bonded interactions in the potential of Eq.~\ref{eq:canonical_potential} takes the following form:
\begin{equation}
\label{vdh}
V_{\rm DH}(T,I) = \sum_{ij} \dfrac{(q_{\rm eff} e)^2}{4\pi \epsilon_0 \epsilon_{\rm r}}
\dfrac{
\exp{\left\{-r_{ij}^{\text{b-b}}/\lambda_{\rm DH}(T,I)\right\}}
}{r_{ij}^{\text{b-b}}} \text{,}
\end{equation}
where $q_{\rm eff}$ is the effective charge situated at the backbone site of each of the nucleotides, 
$r_{ij}^{\text{b-b}}$ is the magnitude of the distance between the backbone
sites of nucleotides $i$ and $j$, $\epsilon_0$ is the permittivity of the
vacuum, $\epsilon_r$ is the relative permittivity of water, $e$ is the
elementary charge and $q_\text{eff}$ is a dimensionless effective charge. In principle, $\epsilon_r$
depends on $r_{ij}$,\cite{mazurDielectric} and weakly depends on
temperature and salt concentration. However, for oxDNA2 we set $\epsilon_r$ to be a constant value, in
keeping with the coarse-grained approach taken for the rest of the model. In particular, we
choose $\epsilon_r = 80$, the standard value for water.  In Eq.~\ref{vdh} we have stressed 
that the interaction depends on the temperature $T$ and on the (monovalent) salt
concentration $I$ through the Debye length $\lambda_{\rm DH}(T,I)$:
\begin{equation}
\label{lambdadh}
\lambda_{\rm DH}(T,I) = \sqrt{\dfrac{\epsilon_0 \epsilon_{\rm r} k_{\rm B}T}{2 N_{\rm A} e^2 I}},
\end{equation}
where $N_{\rm A}$ is Avogadro's number and $k_{\rm B}$ is Boltzmann's constant.

\begin{figure}[tb]
\begin{tikzpicture}
\begin{scope}
\draw node[anchor=north west] (a) {(a)};
\draw ($(a.north) + (-0.25cm,0.0cm)$) node[anchor=north west] (tvq) {\includegraphics{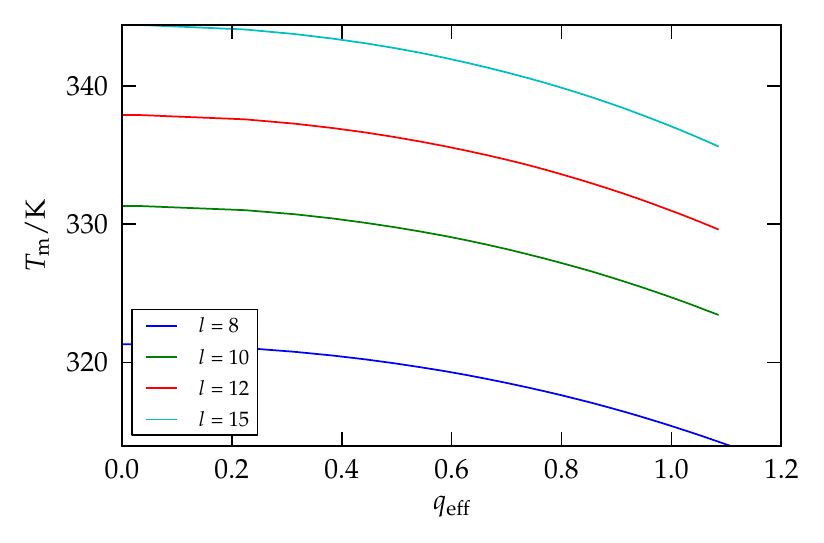}};
\draw ($(a.north west) + (0.0cm,-5.5cm)$) node[anchor=north west] (b) {(b)};
\draw ($(b.north) + (0.0cm,0.0cm)$) node[anchor=north west] (dtmvq) {\includegraphics{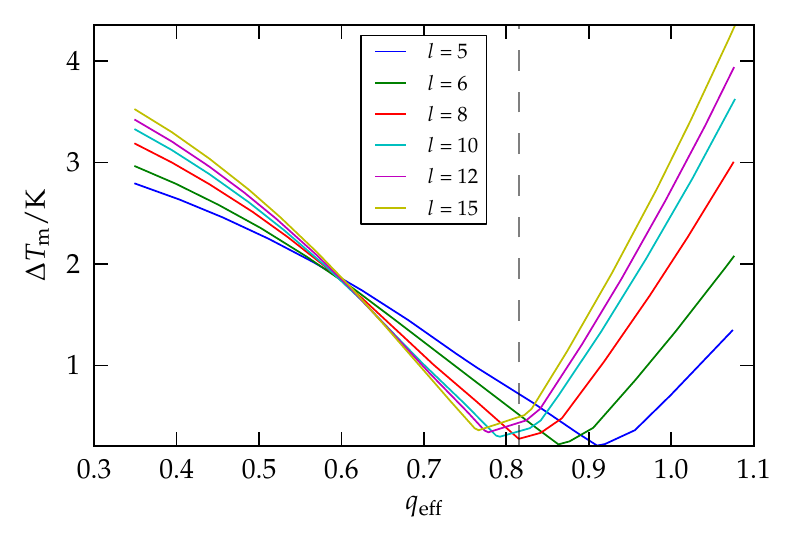}};
\end{scope}
\end{tikzpicture}
\caption{\label{fig:ferd}(a) Melting temperature of duplex
  DNA of different lengths $l$ at [Na$^+$]=0.2\,M as a function of the effective charge
  $q_{\rm eff}$ obtained with thermodynamic integration.
  (b) Average melting temperature difference between our model and the SantaLucia model\cite{SantaLucia2004}
  for different values of $l$ as a function of $q_{\rm eff}$.
  Each point corresponds to the average over [Na$^+$]=0.1, 0.2, 0.3, 0.4, 0.5\,M of the magnitude of the
  difference in melting temperature between our model and the SantaLucia model.
  The plot suggests an optimal value for $q_{\rm eff}$ of $0.815$, indicated by the vertical dashed line.}
\end{figure}

To ensure the computational efficiency of simulating the model, we set the
interaction to zero at a finite distance; and to allow simulation with molecular 
dynamics, we introduce a quadratic smoothing potential so that the interaction goes to
zero smoothly. The quadratic smoothing, the details of which are reported
in Appendix~\ref{appendix:debye huckel}, is introduced after a cutoff
$r_{\rm smooth}$, which we choose to be $3\lambda_{\rm DH}$. This cutoff allows us to use
all the standard techniques to improve the simulation efficiency via the use of
Verlet lists and/or cells.\cite{FrenkelBook} We have checked that introducing our chosen cutoff, $r_{\rm smooth} = 3\lambda_{\rm DH}$, has a negligible effect on the duplex thermodynamics results used to parameterise the interaction (Fig.~\ref{fig:ferd}(a)).

Our representation of DNA uses a single rigid body per nucleotide, and
the best choice of where to put the charge is not obvious. All the atoms of
the sugar and phosphate groups of the backbone are represented in a single
interaction site, and it is thus natural to put the charge, which in real DNA is
located on the phosphate, on that interaction site. Importantly, the backbone
site of a nucleotide is placed almost in between the phosphate of that
nucleotide and the phosphate of the neighbouring one, which could potentially lead to some
unphysical effects. Also, we should stress that having a charge at each backbone
site means that the DNA has as many charges as nucleotides, which is
not always true in real systems: very often, the terminal phosphate at the
$3^\prime$ end is cut off, removing a charge.  The absence of this charge can
cause measurable effects on the thermodynamics, and indeed the SantaLucia model\cite{SantaLucia2004}
requires the presence or absence of the terminal phosphates as an input parameter.
In keeping with our coarse-graining approach, we put a half effective charge on
each of the terminal nucleotides to incorporate the fact that each charge should
be halfway in between our backbone sites, emulating a system with the terminal 
charge removed, and parameterise to the SantaLucia model in a way consistent with this approach.

The parameter that we have tuned to reproduce the thermodynamics predicted by
the model of SantaLucia is the effective charge $q_{\rm eff}$. To do this,
we used thermodynamic integration~\cite{vega2008thermodynamicIntegration} to compute the melting temperatures
of duplexes of length 5, 6, 7, 8, 10, 12 and 15 as a function of $q_{\rm eff}$
at several salt concentrations, and chose the value that best reproduced the
melting temperatures predicted by SantaLucia's model. 

The melting temperature of a duplex is defined as the temperature at which half of the strands in a stoichiometric bulk solution are in the duplex state. 
 We cannot simulate the bulk system with oxDNA. However, at the bulk melting temperature for a given concentration of strands, the bound and unbound free energies obtained for a system of two strands at the same concentration,  $F_{\rm b}$ and $F_{\rm ub}$, are related by:\cite{Ouldridge_bulk_2010}
\begin{equation}
\label{td1}
F_{\rm b}(T_{\rm m},q_{\rm eff}) = F_{\rm ub}(T_{\rm m},q_{\rm eff}) - k_{\rm B}T_{\rm m} \ln(2)\,,
\end{equation}
where we have made explicit the dependence of $F$ on the effective charge $q_{\rm eff}$. The
constant on the right hand side of Eq.~\ref{td1} accounts for concentration
fluctuations that are present in bulk but not for two strands in a periodic box.\cite{Ouldridge_bulk_2010} The equation is exact in the dilute limit, which is indeed where the
thermodynamics of duplex formation are usually studied and is the relevant limit for oxDNA.

We can take advantage of the relation of the free energies given by Eq.~\ref{td1} for the thermodynamic integration. For small changes in $T$ and $q_{\rm eff}$, it is possible to write

\begin{align}
\label{td2}
&F(T_{\rm m} + {\rm d}T, q_{\rm eff} + {\rm d}q_{\rm eff}) = \nonumber\\
&F(T_{\rm m}, q_{\rm eff}) + \dfrac{\partial F}{\partial T} {\rm d}T +
\dfrac{\partial F}{\partial q_{\rm eff}} {\rm d}q_{\rm eff} \nonumber = \\
&F (T_{\rm m}, q_{\rm eff}) + \left(\frac{F-\langle V \rangle}{T_{\rm m}} + \left\langle \frac{\partial V}{\partial T} \right\rangle \right){\rm d}T + 2\dfrac{\langle V_{\rm DH} \rangle}{q_{\rm eff}} {\rm d}q_{\rm eff}.\,
\end{align}
The last step includes the unusual term $\langle \partial V/\partial T \rangle $. This is because 
our potential, like many coarse-grained models, depends explicitly on $T$
through $V_{\rm stack}$ and $V_{\rm DH}$. 

To compute the change in $T_{\rm m}$ introduced by a small change in $q_{\rm eff}$,
one can impose the condition in Eq.~\ref{td1} at the new $T_{\rm m}$
\begin{align}
\label{td3}
F_{\rm b}(T_{\rm m} + {\rm d}T,q_{\rm eff} + {\rm d}q_{\rm eff}) &= \nonumber \\
F_{\rm ub}(T_{\rm m} + {\rm d}T,q_{\rm eff} + {\rm d}q_{\rm eff}) &- k_{\rm B}(T_{\rm m} + {\rm d}T) \ln(2)\,,
\end{align}
and, by using Eqs.~\ref{td1}, \ref{td2} and \ref{td3}, one obtains
\begin{align}
\label{td4}
\dfrac{{\rm d}T_{\rm m}}{{\rm d}q_{\rm eff}} = \dfrac{(2/q_{\rm eff})(\langle V_{\rm
DH}\rangle_{\rm ub} -\langle V_{\rm DH}\rangle_{\rm b})}{(\langle V_{\rm ub} \rangle -\langle V_{\rm b} \rangle)/T_{\rm m} - (\langle \partial V_{\rm ub}/\partial T \rangle -\langle \partial V_{\rm b} /\partial T \rangle)}.
\end{align} 
Eq.~\ref{td4} is a differential equation that allows us to follow the
change in melting temperature as $q_{\rm eff}$ is changed. We note that
Eq.~\ref{td4} is an extended Clausius-Clapeyron relation,\cite{vega2008thermodynamicIntegration}
and the quantities on the right-hand side are readily accessible with separate
simulations of the bound and unbound states.  Since we are dealing with small
systems, all these simulations are very quick and it is thus easy to achieve
very accurate results. As a starting point for the thermodynamic integration we
use the melting temperatures from SantaLucia, which our original model
(equivalent to the current model at $q_{\rm eff} = 0$) reproduces within tenths
of a Kelvin. Since both $V_{\rm stack}$ and $V_{\rm DH}$ only
weakly depend on temperature, we used Eq.~\ref{td4} assuming $\langle \partial V_{\rm ub}/\partial T \rangle -\langle \partial V_{\rm b} /\partial T \rangle=0$ to obtain an optimal value for $q_{\rm eff}$, and checked {\it a posteriori} the results of
thermodynamic integration with melting simulations. We have found that the
melting temperature differences computed with Eq.~\ref{td4} are accurate to within
$0.2$\,K.

For a range of salt concentrations and duplex lengths, we performed
thermodynamic integration to find the melting temperature as a function of
$q_{\rm eff}$. Some of the results are depicted in Fig.~\ref{fig:ferd}. The melting
temperature of the duplex decreases with increasing effective charge $q_{\rm eff}$, but not
dramatically so (Fig.~\ref{fig:ferd}(a)). This is because the increased interchain
repulsion in the duplex state is partially balanced by the lower entropy of the single-stranded
state (due to it adopting a slightly more extended state to reduce intrachain repulsion).
The data in Fig.~\ref{fig:ferd}(b) show that the $q_{\rm eff}$ at which the difference in
melting temperature between oxDNA and SantaLucia is minimised depends on the duplex
length and lies in the range $0.75 < q_{\rm eff} < 0.95$. The best overall predictions are
obtained for an effective charge $q_{\rm eff}=0.815$. It is reassuring that this value does 
not deviate significantly from $1$ (corresponding to the value given by Debye-H{\"u}ckel theory), 
and that the best value for $q_{\rm eff}$ varies little with duplex length. In addition, it is not uncommon to use a value of $q_{\rm eff} < 1$ for coarse-grained DNA models.\cite{3spnSalt, Maffeo2010} One argument is that ion condensation, which is known to occur for DNA, will screen the phosphate charges more strongly than expected from Debye-H{\"u}ckel theory. This will lead to a lower effective charge when fitting a model using a Debye-H\"uckel treatment to experimental results, although such arguments should be applied with caution to a crude mean-field approach such as this one.

We note that introducing this explicit electrostatic term in our model potential will raise the computational expense of each simulation step compared to the original oxDNA model, as the electrostatic term will generally result in oxDNA2 having a larger interaction range than the original model has. This effect increases at lower salt, as the Debye-H\"uckel term becomes more long-ranged. For example, we find that simulating a 10-bp duplex with MD for a given number of steps takes $1.4\times$ as long with oxDNA2 at [Na$^+$]=0.5\,M and $1.7\times$ as long with oxDNA2 at [Na$^+$]=0.1\,M as it did for the original oxDNA.

\section{Improving structure prediction for large-scale DNA objects}
\label{section:global twist param}

\begin{figure}
  \includegraphics[width=0.3\linewidth]{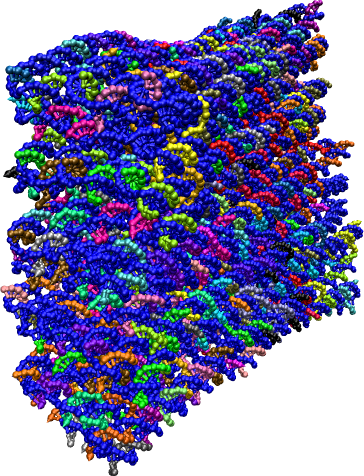}
  \includegraphics[width=0.3\linewidth]{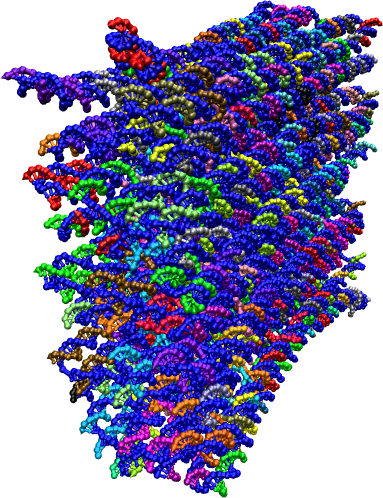}
  \includegraphics[width=0.3\linewidth]{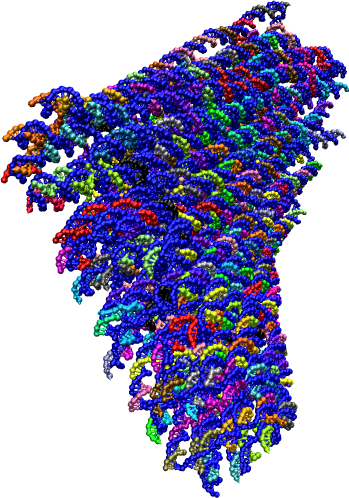}
  \caption{Simulation snapshots for the L- (left), N- (middle), and R- (right) type helix bundles, each composed of roughly 15,000 nucleotides. The designs are taken from Dietz \text{et al}.\cite{3dDietz2009} In experiment and in oxDNA2, the N-type helix bundle exhibits close to zero global twist.}
  \label{fig:monoliths sim}
\end{figure}

The model was originally parameterised for small single- and double-stranded DNA structures. If we wish to study large-scale structures, we need to ensure that we can reproduce the existing experimental data for these larger constructs. A good test case is provided by the work of Dietz \textit{et al.},\cite{3dDietz2009} who measured the global twist of three different DNA origami structures, described as 10-by-6 helix bundles (Fig.~\ref{fig:monoliths sim}). We denote the three origami structures as L-, N- and R-type. In the experiment, they were designed to impose a pitch of 10, 10.5, and 11 base pairs per turn (bp/turn) on the constituent DNA double helices and these different designs exhibited left-handed, no, and right-handed global twist, respectively, when multimerised to form ribbons and visualised with transmission electron microscopy. One might think that this result implies that DNA has a natural pitch of 10.5 bp/turn, since the design with that inherent periodicity did not result in a globally twisted system. However, it is not that straightforward.

\begin{table}
  \begin{tabular}{c c c}
    \hline
    \hline
    & \multicolumn{2}{c}{Global Twist/${}^\circ$} \\
    \cline{2-3}
    Helix Bundle Type & oxDNA2 & experiment \\ \hline
    N & 0.2 & 0 \\ 
    L & -22.2 & -31 +/- 5 \\
    R & 26.9 & 26 +/- 5 \\
    \hline
    \hline
  \end{tabular}
  \caption{Global twist in the DNA helix bundles of Dietz \textit{et al.}\cite{3dDietz2009} for simulations using oxDNA2 and experiment. The oxDNA2 simulations were run with [Na$^+$]=0.5\,M for $1.8\times 10^9$ molecular dynamics (MD) steps for each design.}
  \label{table:monolith twist table}
\end{table}

Unsurprisingly, when simulated with the original version of oxDNA, the N-type origami still showed a significant right-handed global twist, chiefly because the duplex pitch for the model was 10.36 bp/turn. However, even when simulated with a version of oxDNA modified so that the model duplex pitch was 10.5 bp/turn, the N-type origami displayed a right-handed twist. One reason for this is the following: in oxDNA, the helical twist across nicks and junctions is larger by about $3.9^{\circ}$ and $2.5^{\circ}$ degrees respectively than for a normal duplex step (we call this an overtwist), so that fractionally fewer base pairs are required for a helix turn than would otherwise be expected. Although these differences are small, they add up constructively to create a global twist on a structure that would otherwise have no such twist. For a nicked duplex in oxDNA, the overtwist occurs because, opposite the nick, the non-nicked strand prefers a larger twist than for duplex DNA (just as stacked single strands do in the model), and the nicked backbones lack the FENE spring that would usually oppose this tendency. A similar argument applies for strands with junctions. 

Although there is no direct evidence available to show whether this overtwist is physically realistic, there are multiple lines of experimental evidence that suggest that the pitch of duplex DNA is close to 10.5 bp/turn.\cite{taylor1990cyclisation,shore1983} Given the evidence from Dietz \textit{et al.}\cite{3dDietz2009} that the effective pitch in DNA origami structures is also close to 10.5 bp/turn, we decided to reduce the overtwisting in oxDNA as much as is possible. To achieve this, we modify the coaxial stacking term of the potential,  $V_\text{coax stack}$, so that the overtwist is $0^{\circ}$ for a junction and $1.3^{\circ}$ for a nick (see Appendix \ref{appendix:coax change} for details of the changes to the potential). We have verified that this change has very little effect on the other features of the model.

Even when the overtwist at nicks and junctions has largely been removed, we require an intrinsic duplex pitch of $10.55$ bp/turn in order for the oxDNA N-type helix bundle to have zero global twist, and we set the pitch to this value (at [Na$^+$]=0.5\,M) for oxDNA2 by modifying $V_\text{backbone}$. We found that this modification to the backbone potential changed the duplex melting temperatures by around 2\,K, and we refitted the model's hydrogen bonding and stacking strengths to correct for this using the same method as was used for the thermodynamics refitting when implementing unequal helical groove widths. The requirement for a pitch of $10.55$ bp/turn, rather than the $10.5$ bp/turn that one might expect, is due to a few subtle effects that are currently being investigated further.

The global twist measured for the helix bundles with the new oxDNA2 model (Appendix~\ref{appendix:global twist} describes how the global twist was measured from simulations) is compared to the experimental results in Table~\ref{table:monolith twist table}, while typical simulation snapshots are shown in Fig.~\ref{fig:monoliths sim}. The slight modifications to $V_\text{backbone}$ used to set the model pitch are given in Appendix~\ref{appendix:pitch modification}. We note that the modifications to $V_\text{backbone}$ changed the duplex melting temperatures in the model by 1-2\,K, which were then refitted using histogram reweighting (as described in Section II C 2 of Ref.~\onlinecite{necitujoxRNA}) to give an agreement with experimental melting temperatures that was as good as for the original model.

\section{AA/TT sequence dependence}
\label{section:aa/tt}

As a further improvement, the oxDNA2 model incorporates a more realistic sequence-dependent stacking interaction, which is achieved by differentiating between the AA and TT stacking interaction strengths. In the previous parameterisation of oxDNA,\cite{petrSeqDep2012} the sequence-dependent base-pairing and stacking interaction strengths were obtained by fitting the oxDNA duplex melting temperatures to the SantaLucia model, a nearest-neighbor model that is able to predict experimental duplex melting temperatures very well.\cite{SantaLucia2004} The SantaLucia model is designed at the level of base-pair steps, where each base-pair step consists of four bases, with a free-energy difference between its single-stranded state and its duplex state. In total there are 10 unique base-pair steps in the SantaLucia model: AA/TT, AT/AT, TA/TA, GC/GC, CG/CG, GG/CC, GA/TC, AG/CT, TG/CA, GT/AC (for example, AG/CT refers to a base-pair step with complementary bases AG on one strand and CT on the other, both specified in the $3^\prime$ to $5^\prime$ direction).  Therefore, fitting oxDNA to the SantaLucia model only allows one to find the sum of the strengths of the two stacking interactions between the nucleotides that are within a base-pair step. For example, it is only possible to find the average of the AA and TT stacking strengths from the SantaLucia model. 

However, experimental evidence from sequence-dependent measurements of the mechanical properties of single-stranded DNA,\cite{plaxco_ssDNA, ritort2013ssDNAsalt} as well as hairpin stabilities and closing rates, \cite{libchaber_prl2000, libchaber_pnas} have revealed that sequences of A bases are much stiffer than equal length sequences of T bases. It has been argued that this is evidence that consecutive AA bases stack much more strongly than TT bases do. Here we use original experimental data on the stabilities of hairpins with either a poly-A or a poly-T loop, to differentiate between the AA and TT stacking strengths. All hairpins have 6-bp stems and loops of either $21$ or $31$ bases. Details of the experimental systems and the sequences are given in Appendix~\ref{appendix:experimental hairpin}. 

To parameterise our model to reproduce the experimental data, our procedure is to vary the AA stacking interaction strength, $\epsilon_{\rm AA}$, while fixing the sum of the AA and TT stacking interaction strengths to the value that is obtained by fitting oxDNA's duplex melting temperatures to the SantaLucia model, i.e.
\begin{equation}
 \label{eq:eAA}
 \epsilon_{\rm AA} + \epsilon_{\rm TT} = 2 \epsilon_{\rm avg}\text{,}
\end{equation}
where $\epsilon_{\rm TT}$ is the TT stacking strength, $\epsilon_{\rm avg}$ is the strength for both AA and TT obtained from fitting oxDNA to the SantaLucia model.
We vary the AA stacking interaction strength to match the experimental differences of the thermal stabilities of the poly-A-loop and poly-T-loop hairpins, specifically $\delta \Delta F = \Delta F_{\rm (A-loop)} - \Delta F_{\rm (T-loop)}$, where $\Delta F_{\rm (X-loop)} = F_{\rm b, (X-loop)} - F_{\rm ub, (X-loop)}$ is the difference in free energy between the bound (hairpin) state and the unbound (open) state for a hairpin with a poly-X loop.

\begin{figure}
  \centering
  \includegraphics{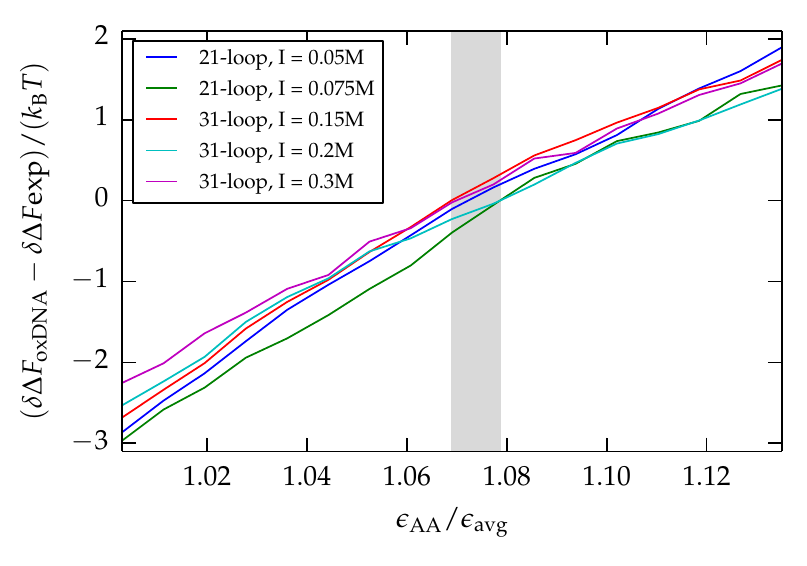}
 \caption{Deviation of oxDNA's value for $\delta \Delta F$ from experiment at 295.6K, as a function of $\epsilon_{\rm AA} / \epsilon_{\rm avg}$. The grey region contains the zero-deviation points for every curve. The experimental results are shown in Fig.~\ref{fig:experimental hairpin deltaF}, and details of the experimental setup can be found in Appendix~\ref{appendix:experimental hairpin}.}
 \label{fig:AA_TT}
\end{figure}

We obtain the bound and unbound free energies $F_{\alpha}$, where ${\alpha} \in \{{\rm b, ub}\}$, by using thermodynamic integration. We start by calculating $F_{\alpha}^{(0)}$, the free energy of a reference state for which the effective charge on the backbone site, $q_{\rm eff}$, is set to 0, and $\epsilon_{\rm AA}=\epsilon_{\rm TT}=\epsilon_{\rm avg}$. $F_{\alpha}(I, x, q_{\rm eff})$, the free energy of the state with salt concentration $I$, stacking strengths $\epsilon_{\rm AA}$ and $\epsilon_{\rm TT}$, and effective backbone charge $q_{\rm eff}$, is then obtained by solving the following integrals
\begin{align}
\label{eq:TI1}
F_{\alpha} (I, x, q_{\rm eff})  =  F_{\alpha}^{(0)} 
 &+ \int_0^{q_{\rm eff}} {\rm d}q_{\rm eff}^{\prime}  \frac{\partial F_\alpha(I,0,q_{\rm eff}^{\prime})}{\partial q_{\rm eff}^{\prime}} \\ \nonumber
 &+ \int_0^x {\rm d}x^\prime  \frac{\partial F_\alpha(I,x^{\prime},q_{\rm eff})}{\partial x^{\prime}} 
\end{align}
where $x$ measures the deviation of the AA and TT strengths from $\epsilon_{\rm avg}$, such that Eq.~\ref{eq:eAA} is satisfied, i.e.
\begin{equation}
  \label{eq:x def}
 \epsilon_{\rm AA} = (1+x)\epsilon_{\rm avg}, ~~ \epsilon_{\rm TT} = (1-x) \epsilon_{\rm avg}\,\text{.} 
\end{equation}
Taking the derivatives of the free energy in Eq.~\ref{eq:TI1} we obtain,
 \begin{align}
 \label{eq:TI2}
 F_{\alpha} (I, x, q_{\rm eff}) =& F_{\alpha}^{(0)} \nonumber + \int_0^{q_{\rm eff}} {\rm d}q_{\rm eff}^{\prime}  \frac{2\langle V_{\rm DH}(I,0,q_{\rm eff}^{\prime})\rangle_{\alpha}}{q_{\rm eff}^{\prime}} \nonumber \\
&+ \int_0^x {\rm d}x^{\prime}   \left[  \frac{1}{{1+x^{\prime}}} \langle V_{\rm stack}(I,x^{\prime},q_{\rm eff}) \rangle_{\alpha}^{\rm AA} \right. \nonumber \\
& \left. - \frac{1}{{1-x^\prime}} \langle V_{\rm stack}(I,x^{\prime},q_{\rm eff}) \rangle_{\alpha}^{\rm TT}     \right] \,\text{,}   
 \end{align}
where $ \langle V_{\rm \small stack}(I,x,q_{\rm eff}) \rangle_{\alpha}^{\rm \small AA(TT)}$ is the average stacking energy of all AA(TT) nucleotides in a strand in state $\alpha$ with salt concentration $I$, $x$ defined as in Eq.~\ref{eq:x def}, and backbone charge $q_{\rm eff}$. The terms appearing in the integrands of Eq.~\ref{eq:TI2} can be obtained by running short simulations of bound and unbound states separately and therefore the free energies can be calculated in an efficient manner. The results for the difference in $\delta \Delta F$ between oxDNA and experiment as a function of $\epsilon_{\rm AA}$ for the hairpins with 21- and 31-base loops are shown in Fig.~\ref{fig:AA_TT}; the value for the AA stacking strength that minimises this difference is found to be $\epsilon_{\rm AA}\approx 1.075 \epsilon_{\rm avg}$ (corresponding to $\epsilon_{\rm TT}\approx 0.925\epsilon_{\rm avg}$), which is not too dissimilar to the preliminary value suggested in Ref.~\onlinecite{petrSeqDep2012}. We note that this value gives satisfactory predictions for $\delta \Delta F$ for a wide range of salt concentrations down to 0.05\,M.

\section{The oxDNA2 model}
\label{section:oxDNA2 summary}

In summary, the oxDNA2 potential can be written as

\begin{equation}
\begin{split}
V_{\rm oxDNA2} = \sum_{\text{nearest neighbours}}(V_{\text{backbone}}^* + V_{\text{stack}}^* + V'_{\text{exc}}) \;+ \\
\sum_{\text{other pairs}}(V_{\text{HB}}^* + V_{\text{cross stack}} + V_{\text{exc}} + V_{\text{coax stack}}^* + V_{\text{DH}}^*) \text{,}
\end{split}
\label{eq:oxDNA2_potential}
\end{equation}
where a $V_{\rm x}^*$ indicates that the term is either modified for oxDNA2, or, in the case of $V_{\rm DH}^*$, new in oxDNA2. The modified parameters for oxDNA2 are compared with those for oxDNA in Table \ref{table:oxDNA2 params}, and a full account of the changes is given in Appendix \ref{appendix:oxdna2 potential}. All other parameters remain the same as in the original model. 

We emphasise that, after all the relevant changes to the potential were made, the hydrogen bonding and stacking parameters were modified to ensure that the close agreement to experimental duplex melting temperatures achieved for the original model was retained with oxDNA2 (this was done immediately after the changes to $V_{\rm backbone}$, as described at the end of Section \ref{section:global twist param}).

\section{Physical properties of oxDNA2}
\label{section:physical properties}
The structural, mechanical and thermodynamic properties of DNA for the new version of oxDNA presented in this paper, which we call oxDNA2, are slightly different from the properties for the original oxDNA model. We briefly highlight the most important of these changes here; in addition these properties are given as a function of salt concentration as this is now possible with the new model. We note that all of the results described in this section were computed using the final version of the oxDNA2 model as summarized in Section \ref{section:oxDNA2 summary}. The details of the simulations used to compute these results are given briefly in the figure captions and in detail in Appendix~\ref{appendix:simulation details} and Table~\ref{table:simulation parameters}.

\begin{figure}
\begin{tikzpicture}
\begin{scope}
\draw node[anchor=north west] (a) {(a)};
\draw ($(a.north east) + (-0.7cm,0.0cm)$) node[anchor=north west] (rise) {\includegraphics{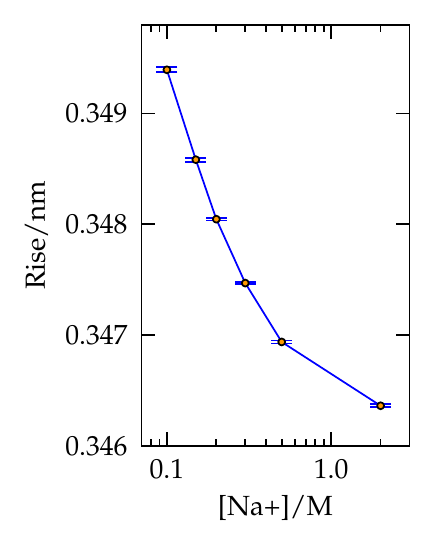}};
\draw ($(rise.north east) + (-0.3cm,0.0cm)$) node[anchor=north west] (pitch) {\includegraphics{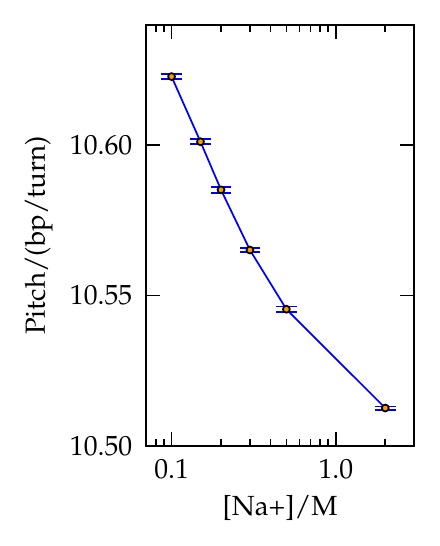}};
\draw ($(rise.north east) + (-0.2cm,0.0cm)$) node[anchor=north west] (b) {(b)};
\end{scope}
\end{tikzpicture}
  \caption{(a) Rise and (b) pitch of a 60-bp duplex as a function of salt concentration in oxDNA2. For each salt concentration the duplex was simulated for at least $3\times10^9$ MD steps. The error bars, which are narrower than the plot markers, show the standard error on the mean given by averaging over 10 independent estimates for each data point.}
  \label{fig:pitch and rise}
\end{figure}

The structural properties of the new model, specifically the pitch and rise of double-stranded DNA, are presented in Fig.~\ref{fig:pitch and rise}. As might be expected, the rise increases with decreasing salt concentration, due to the greater repulsion between backbone sites. The pitch also increases with decreasing salt, consistent with the measured increase in rise and slight decrease in neighbouring backbone-backbone distance measured as salt concentration is decreased (the duplex radius remains approximately constant). Although there is not much experimental evidence to compare this with, there is some indication that the pitch is roughly constant for low salt concentrations (0.162\,M and below).\cite{taylor1990cyclisation} As mentioned earlier, the pitch is chosen (by modifying the bonded neighbour backbone-backbone interaction) so that the global twist of origami structures agrees with experimental measurements. Specifically, we set the backbone-backbone interaction so that the helix bundle designed to have no global twist has no global twist in the model at [Na$^+$]=0.5\,M. This results in a pitch of roughly $10.55$ bp/turn at this salt concentration, compared to $10.36$ bp/turn in the original model, and experimental values of around 10.45 suggested by cyclisation experiments,\cite{taylor1990cyclisation,shore1983} albeit in the presence of some divalent salt.

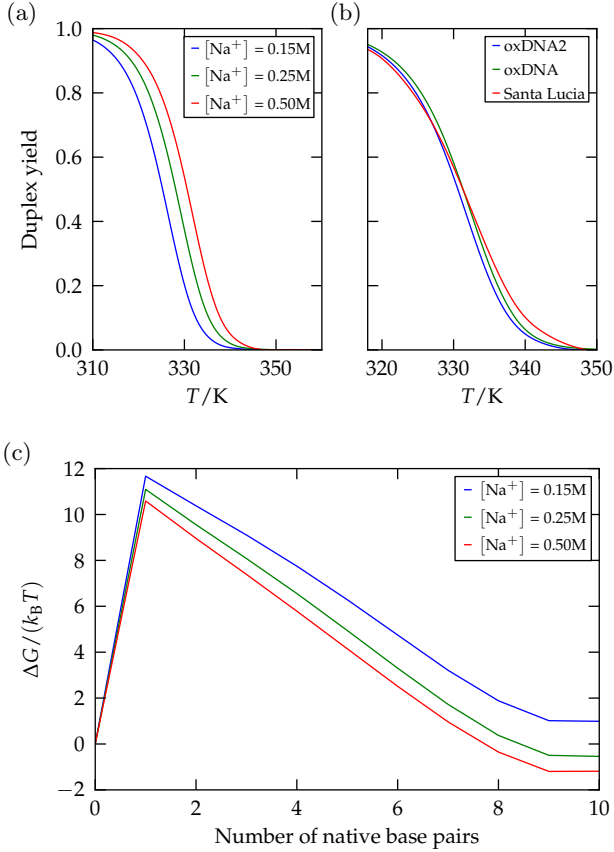
\begin{figure}
\begin{tikzpicture}
\begin{scope}
\draw node[anchor=north west] (a) {(a)};
\draw ($(a.north east) + (-0.7cm,0.0cm)$) node[anchor=north west] (model) {\includegraphics{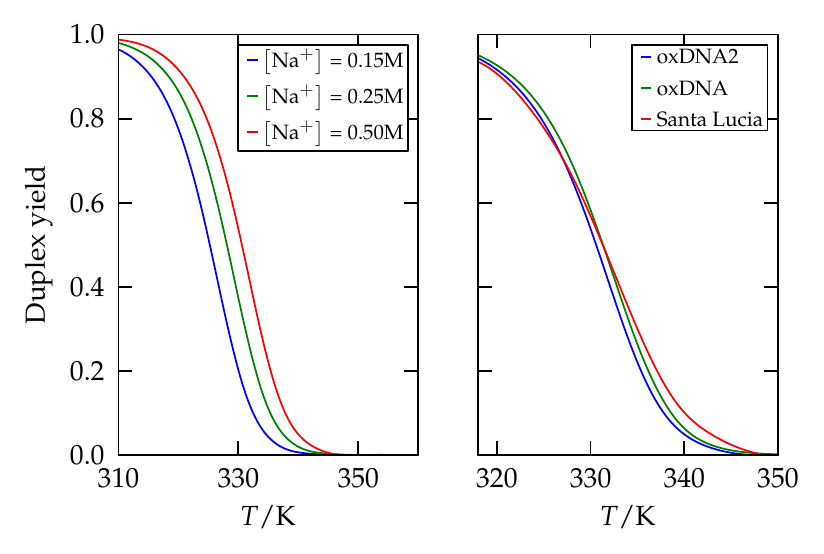}};
\draw ($(a.north east) + (3.65cm,0.0cm)$) node[anchor=north west] (b) {(b)};
\draw (model.south west) node[anchor=north west] (c) {(c)};
\draw (model.south west) node[anchor=north west] {\includegraphics{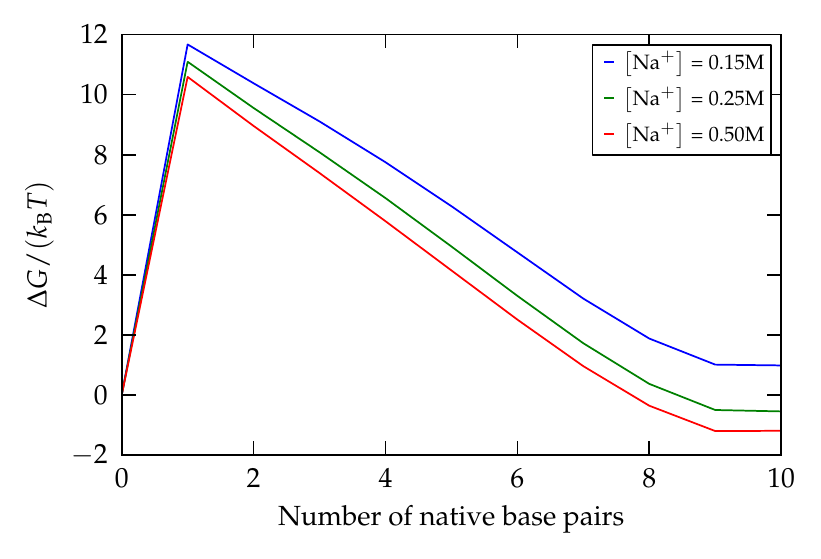}};
\end{scope}
\end{tikzpicture}
  \caption{The thermodynamics of formation of a 10-bp duplex with oxDNA2. (a) Duplex yield for oxDNA2 at different salt concentrations. (b) Comparison of duplex yield profiles at [Na$^+$]=0.5\,M as predicted by oxDNA2, the original oxDNA model and the SantaLucia model.\cite{SantaLucia2004} (c) The free-energy profile for duplex formation at different salt concentrations, at the melting temperature of a 10-bp duplex at [Na$^+$]=0.25\,M. The duplex was simulated for at least $3\times 10^{10}$ virtual move Monte Carlo\cite{Whitelam2009} (VMMC) steps at each salt concentration using umbrella sampling at a monomer concentration of $3.3\times 10^{-4}$\,M.}
  \label{fig:duplex thermo}
\end{figure}

The thermodynamics of duplex formation are shown in Fig.~\ref{fig:duplex thermo}. The transition width for the yield curve of a 10-bp duplex at 0.5\,M [Na$^+$] in oxDNA2 is largely unchanged from the original oxDNA, and the transition widths depend weakly if at all on salt in oxDNA2. The free-energy profiles for duplex formation in oxDNA2 show that the free-energy cost of forming the first base pair decreases with increasing salt, presumably due to the reduced energetic cost of bringing the two single strands close together as the electrostatic repulsion between backbone sites becomes more short-ranged. The slope of the bound region of the free-energy profile also decreases with increasing salt, indicating a reduced free-energy cost of forming subsequent base pairs.

\begin{figure}
\begin{tikzpicture}
\begin{scope}
\draw node[anchor=north west] (plot) {\includegraphics{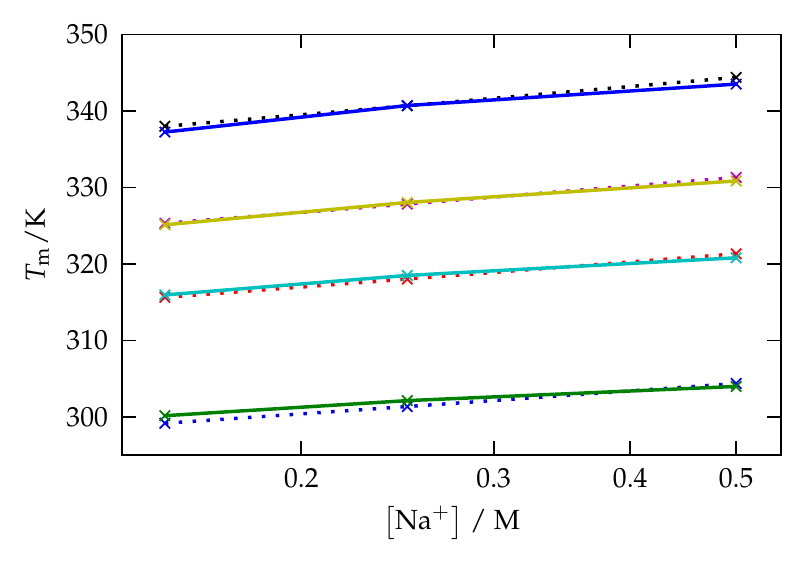}};
\draw ($(plot.north west) + (3.5cm,-0.63cm)$) node[anchor=north west] (15bp) {15-bp duplex};
\draw ($(plot.north west) + (3.5cm,-1.6cm)$) node[anchor=north west] (10bp) {10-bp duplex};
\draw ($(plot.north west) + (3.6cm,-2.4cm)$) node[anchor=north west] (8bp) {8-bp duplex};
\draw ($(plot.north west) + (3.6cm,-3.7cm)$) node[anchor=north west] (6bp) {6-bp duplex};
\end{scope}
\end{tikzpicture}
  \caption{The melting temperature as a function of salt concentration for duplexes of different lengths simulated with oxDNA2 (crosses with solid lines) and as predicted by the SantaLucia model (crosses with dotted lines) for a DNA strand concentration of $3.3\times 10^{-4}$\,M. Each duplex was simulated for roughly $4\times 10^9$ VMMC steps at a salt concentration of $0.25\,$M, and the melting temperatures at $0.15\,$M and $0.5\,$M were computed by single histogram reweighting.\cite{Kumar1992}}
  \label{fig:duplex melting}
\end{figure}
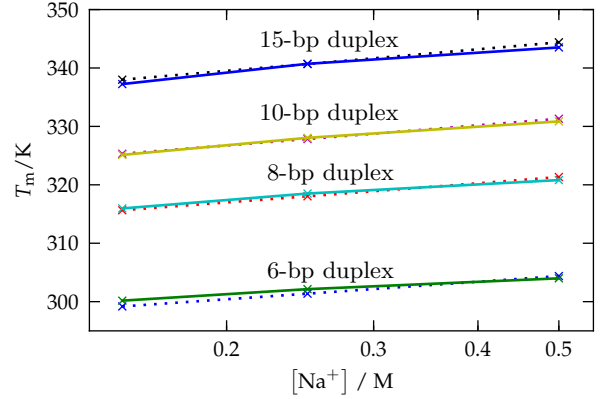

In Fig.~\ref{fig:duplex melting}, the duplex melting temperatures as a function of salt and for different duplex lengths in oxDNA2 are compared to the results from the SantaLucia model, to which oxDNA2 was parameterised. As expected, we find good agreement.

\begin{figure}
\begin{tikzpicture}
\begin{scope}
\draw node[anchor=north west] (a) {(a)};
\draw ($(a.north east) + (-0.7cm,0.0cm)$) node[anchor=north west] (pl) {\includegraphics{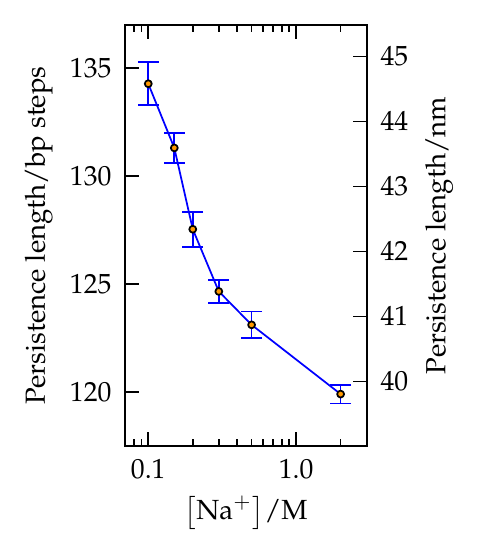}};
\draw ($(pl.north east) + (-0.3cm,0.0cm)$) node[anchor=north west] (ts) {\includegraphics{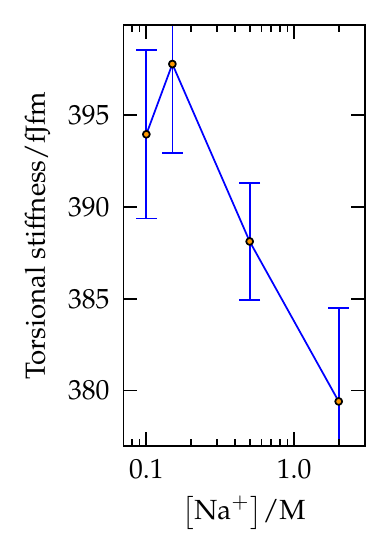}};
\draw ($(pl.north east) + (-0.2cm,0.0cm)$) node[anchor=north west] (b) {(b)};
\end{scope}
\end{tikzpicture}
  \caption{(a) The persistence length and (b) the torsional stiffness of duplex DNA in oxDNA2 as a function of salt concentration. For the persistence length, a 60-bp duplex was simulated for at least $3\times 10^9$ MD steps for each salt concentration. The method for computing the torsional stiffness is given in Appendix~\ref{appendix:torsional stiffness}. Error bars for the simulation results show the standard error on the mean given by averaging over either 10 independent estimates (for the persistence length) or 5 independent estimates (for the torsional stiffness) for each data point.}
  \label{fig:mechanical properties}
\end{figure}

The mechanical properties of the model double-stranded DNA as a function of salt, specifically the persistence length and torsional stiffness, are shown in Fig.~\ref{fig:mechanical properties}. The persistence length was calculated by computing the correlation of the helix axis as described in Ref.~\onlinecite{dnaModelStruct}. The effective torsional stiffness, $C_{\rm eff}$, was computed from the linear regime of the torque response curve of a 60-bp duplex under tension (see Appendix~\ref{appendix:torsional stiffness} for details). The simulations were carried out under a linear force of 30$\,$pN, a high force regime where we expect that $C_{\rm eff}$ approximates the true torsional stiffness $C_0$.\cite{morozTwisting1997,matekPlectoneme} The persistence length at [Na$^+$]=0.5$\,$M, about 123$\,$bp, is very slightly lower than for the original model, 125$\,$bp, and the persistence length at different salt concentrations is consistent with the rather broad range of values suggested by experiment.\cite{savelyev2012,herrerogalanMechanicalIdentities} The decrease in persistence length with increasing salt is expected due to the decrease in repulsion between the duplex's backbone sites, which makes the duplex less stiff, in agreement with experiment.\cite{savelyev2012} The slight decrease in torsional stiffness with increasing salt can be rationalised in the same way. Regardless of the salt concentration, the torsional stiffness measured for oxDNA2 ($C_{\rm eff} \approx 380-400$\,fJfm) is lower than for the original model\cite{matekPlectoneme} ($C_0 \approx 473$\,fJfm). Single-molecule twisting experiments give a value of $C_{\rm eff} \approx 410$\,fJfm for a pulling force of 3.5\,pN\cite{janssenTorque} and a pulling force of 45\,pN,\cite{Bryant2003} at around [Na$^+$]=0.1\,M. There is limited experimental data on the salt dependence of torsional stiffness in DNA; however the torsional stiffness has been reported to be roughly constant in the range [Na$^+$]=0-0.162\,M.\cite{taylor1990cyclisation}

The effects of various motifs on duplex melting temperatures in oxDNA2 are compared to results from the original oxDNA and the SantaLucia model in Table~\ref{table:other motifs}. These effects are either barely changed in oxDNA2, or are now closer to the SantaLucia values than the original oxDNA values were.

\begin{figure}
\begin{tikzpicture}
\begin{scope}
\draw node[anchor=north west] (plot) {\includegraphics{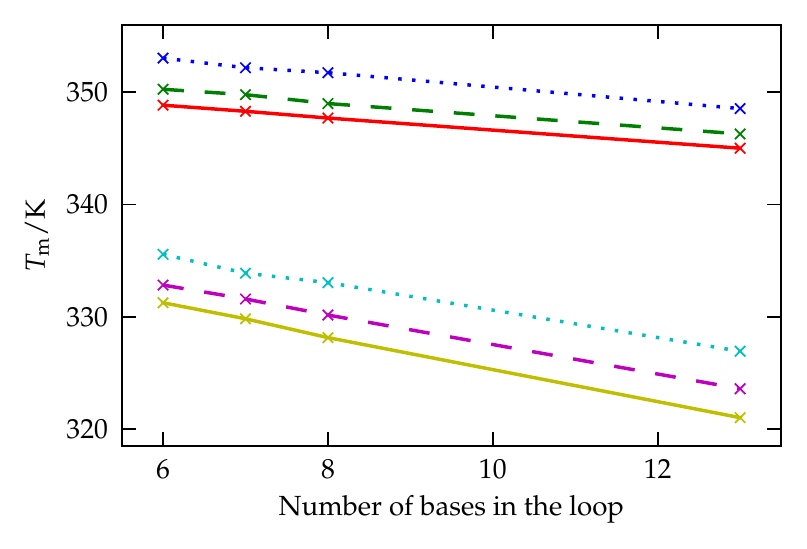}};
\draw ($(plot.north west) + (4.5cm,-0.5cm)$) node[anchor=north west] (12bp) {12-bp stem};
\draw ($(plot.north west) + (3.5cm,-2.5cm)$) node[anchor=north west] (6bp) {6-bp stem};
\end{scope}
\end{tikzpicture}
  \caption{The melting temperature as a function of loop length for hairpins simulated with oxDNA2 (crosses with solid lines), the original oxDNA (crosses with dashed lines), and according to the SantaLucia model (crosses with dotted lines), at 0.5\,M [Na$^+$]. For the oxDNA2 results, the hairpins were simulated at each salt for at least $1.8\times 10^{10}$ VMMC steps.}
  \label{fig:hairpins}
\end{figure}

\begin{figure}
\begin{tikzpicture}
\begin{scope}
\draw node[anchor=north west] (plot) {\includegraphics{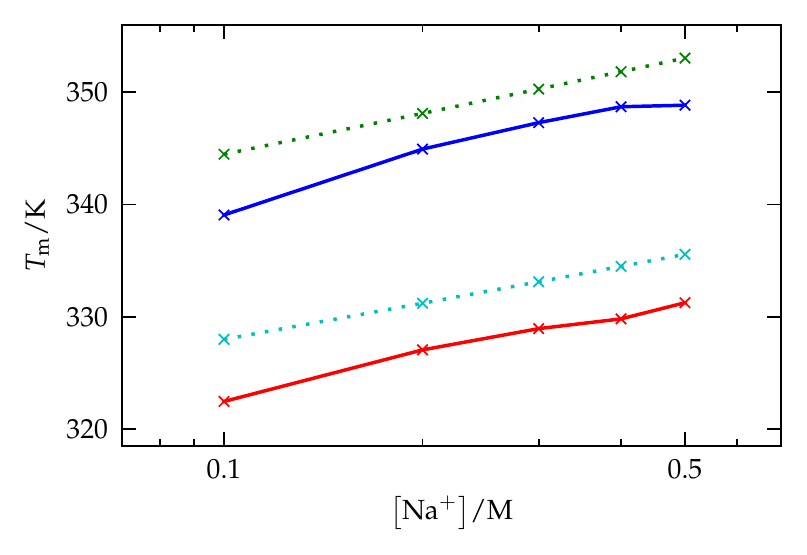}};
\draw ($(plot.north west) + (2.5cm,-0.8cm)$) node[anchor=north west] (12bp) {12-bp stem};
\draw ($(plot.north west) + (4.6cm,-3.7cm)$) node[anchor=north west] (6bp) {6-bp stem};
\end{scope}
\end{tikzpicture}
  \caption{The melting temperature as a function of salt concentration of hairpins with 6-base loops simulated with oxDNA2 (crosses with solid lines) and according to the SantaLucia model (crosses with dotted lines). For the oxDNA2 results, the hairpins were simulated at each salt for at least $1.5 \times 10^{10}$ VMMC steps.}
  \label{fig:low salt hairpins}
\end{figure}

The melting temperatures of hairpins at 0.5\,M [Na$^+$] are shown in Fig.~\ref{fig:hairpins}. The hairpin melting temperatures are lower than for the original oxDNA by about 2\,K, which in turn had hairpin melting temperatures about 3\,K lower than those predicted by SantaLucia. The gap between the oxDNA2 hairpin melting temperatures and those given by the SantaLucia model widens somewhat as the salt concentration is lowered (Fig.~\ref{fig:low salt hairpins}), indicating a stronger salt dependence in oxDNA2, with a typical underestimate of around 7$\,$K at a salt concentration of 100$\,$mM for a relatively short loop length of 6 bases. For longer loops, the difference in predicted melting temperatures between oxDNA2 and SantaLucia widens further. This difference is unsurprising, as in oxDNA2 (and physical DNA) ssDNA becomes stiffer at lower salt concentrations, making the formation of a hairpin less favorable,\cite{majidLoopStacking} whereas in the SantaLucia model the loops' contribution to hairpin stability is salt-independent.\cite{SantaLucia2004}

\begin{figure}
\begin{tikzpicture}
\begin{scope}
\draw node[anchor=north west] (plot) {\includegraphics{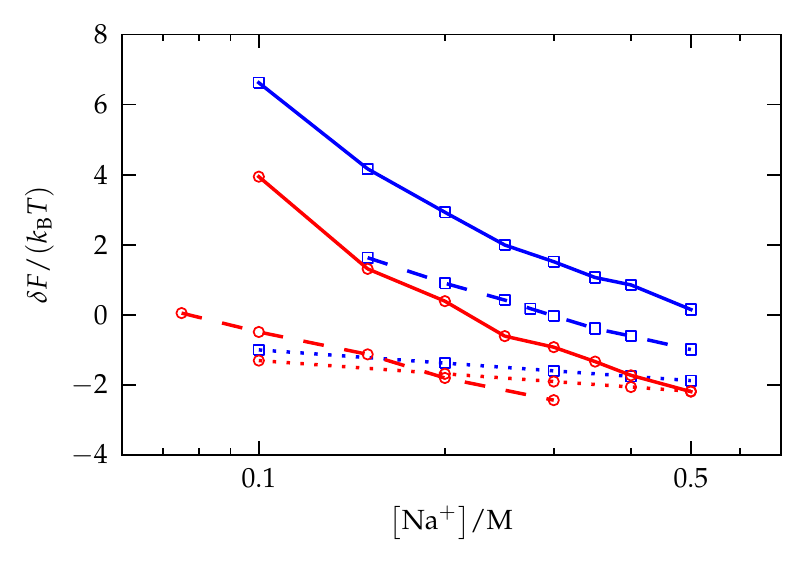}};
\end{scope}
\end{tikzpicture}
  \caption{The free energy difference between the bound and unbound state for two hairpins as measured in experiment using FRET (dashed lines), and as predicted by oxDNA2 (solid lines) and the SantaLucia model (dotted lines). Results for two different hairpins are presented: the $\rm A_{31}\text{-}VW$ (red, circular markers) hairpin and the $\rm T_{31}\text{-}VW$ hairpin (blue, square markers). The sequences for these hairpins are given in Table~\ref{table:experiment hairpin sequence}. The oxDNA2 simulations were run for $10^{10}$ VMMC steps for each hairpin, with $q_{\rm eff}=0$ and $x=0$ (with $x$ defined in Eq.~\ref{eq:x def}), and the results (computed for the standard oxDNA2 values for $q_{\rm eff}$ and $x$) were calculated using thermodynamic integration as described in Section~\ref{section:aa/tt} and Appendix~\ref{appendix:simulation details}.}
  \label{fig:hairpins df performance}
\end{figure}

It seems plausible that oxDNA2's performance is better than implied by the salt-independent loop contribution in the SantaLucia model. In Fig.~\ref{fig:hairpins df performance} we consider hairpins with a short 5-bp stem and a long 31-base loop, making the hairpin thermodynamics particularly sensitive to the change in loop stiffness with salt. The hairpin stability (i.e.~the free energy difference between the bound and unbound state) as predicted by oxDNA2 and the SantaLucia model is compared to experimental data determined using FRET (experimental details in Appendix~\ref{appendix:experimental hairpin}). As expected we find the hairpin stabilities for oxDNA2 have a much stronger salt dependence (steeper gradient) than those of SantaLucia (shallow gradient). At higher salt the oxDNA2 results show a similar slope to the experimental curve, but at lower salt (0.2\,M and below) oxDNA2 shows a stronger destabilization of the hairpins with decreasing salt (a steeper gradient) than experiment. Thus, although oxDNA2 does a reasonable job of capturing these physical effects, the comparison suggests that the single strands may be experiencing too much repulsion at low salt. We also note that the SantaLucia model is unable to predict the large difference in stability of the two hairpins due to their different loop sequences, because the SantaLucia model is insensitive to the loop sequence except for the two bases adjacent to the stem.

\begin{table*}
  \begin{tabular}{c c c c}
    \hline
    \hline
    & \multicolumn{3}{c}{$\Delta T_{\rm m}/\rm K$} \\
    \cline{2-4}
    Motif &  oxDNA2 & oxDNA & SantaLucia \\ \hline
    Bulge (1-base bulge in 8-bp duplex) & $-18.7$ & $-17.98$ & $-23.19$ \\
    Bulge (2-base bulge in 8-bp duplex) & $-27.4$ & $-23.92$ & $-26.73$ \\
    Terminal mismatch (1-bp mismatch in 5-bp duplex) & $+6.5$ & $+6.71$ & $+8.6$ \\
    Internal mismatch (2-bp mismatch in 8-bp duplex) & $-15.9$ & $-15.77$ & $-13.99$ \\
    \hline
    \hline
 \end{tabular}
  \caption{The effect of introducing various motifs on the duplex melting temperature at a monomer concentration of $3.3\times 10^{-4}$\,M. $\Delta T_{\rm m}$ is the difference between the melting temperature of the structure with the motif and a duplex consisting of the same number of complementary base pairs as the motif structure. For bulges and internal mismatches, the motif was placed at the centre of the duplex. The simulations of the bulges, terminal mismatches and internal mismatches were run for at least $6\times 10^{10}$, $8.5\times 10^{10}$, and $5.6\times 10^{10}$ VMMC steps respectively. All oxDNA2 results were obtained from simulations at a concentration of $4.2\times 10^{-5}$\,M which were extrapolated to a concentration of $3.3\times 10^{-4}$\,M.}
  \label{table:other motifs}
\end{table*}

\begin{figure}
\begin{tikzpicture}
\begin{scope}
\draw node[anchor=north west] (a) {(a)};
\draw ($(a.north east) + (-0.7cm,0.0cm)$) node[anchor=north west] (polyA) {\includegraphics{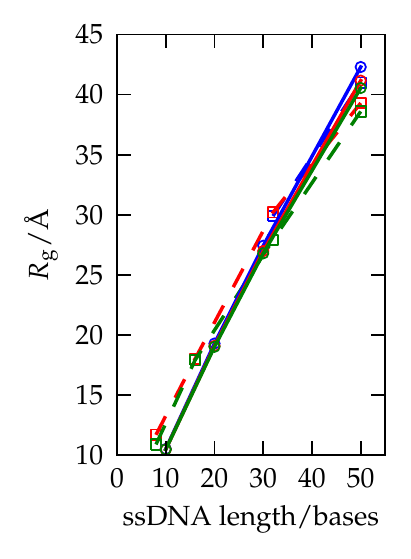}};
\draw ($(polyA.north east) + (-0.3cm,0.0cm)$) node[anchor=north west] (polyT) {\includegraphics{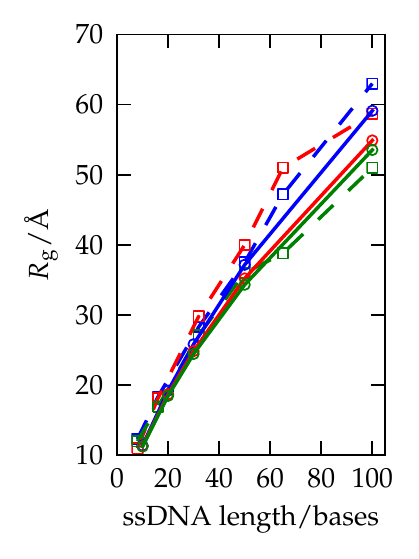}};
\draw ($(polyA.north east) + (-0.2cm,0.0cm)$) node[anchor=north west] (b) {(b)};
\draw ($(polyA.north) + (0.5cm, -0.35cm)$) node[anchor = south] (polyA_label) {poly-A};
\draw ($(polyT.north) + (0.5cm, -0.35cm)$) node[anchor = south] (polyT_label) {poly-T};
\end{scope}
\end{tikzpicture}
  \caption{The radius of gyration ($R_{\rm g}$) for ssDNA as a function of ssDNA length and for different salt concentrations, for (a) a poly-A ssDNA and (b) a poly-T ssDNA, from oxDNA2 (solid lines with circles) and from experiments due to Sim \textit{et al.}\cite{simRg} (dashed lines with squares). The salt concentrations shown are 0.225\,M (blue lines), 0.525\,M (red lines), 1.025\,M (green lines). For the oxDNA2 results, the strands were simulated for between $2\times 10^9$ and $5\times 10^{10}$ VMMC steps at each strand length and salt concentration.}
  \label{fig:rg}
\end{figure}
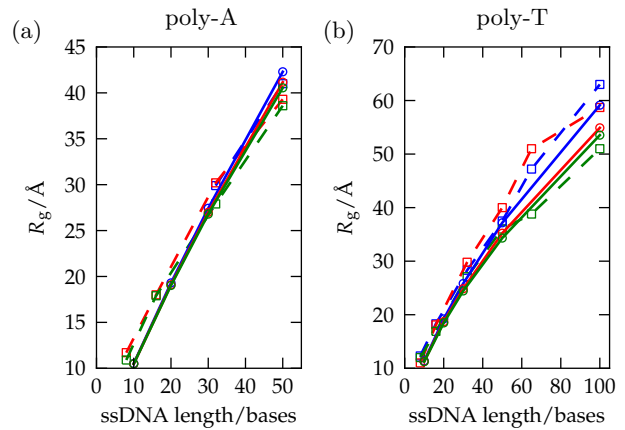

We examine the flexibility of single strands as a function of salt in oxDNA2 by measuring the radius of gyration, $R_{\rm g}$, of a single strand of DNA at different salt concentrations, sequences and strand lengths, and comparing to experimental results (Fig.~\ref{fig:rg}). The oxDNA2 model reproduces the overall trend of increasing $R_{\rm g}$ with salt, and is in agreement with the somewhat noisy experimental data. We note that oxDNA2 is able to capture two important effects: the greater stiffness (and hence larger $R_{\rm g}$) of the more strongly-stacking poly-A strands; and the greater salt dependence of the poly-T strands' $R_{\rm g}$ compared to that of the poly-A strands, caused by the weaker stacking of the poly-T strands, which means that electrostatic repulsion makes a greater relative contribution to its stiffness.

\section{Conclusion}
OxDNA has been applied to a wider variety of systems than any other coarse-grained model of DNA. The modifications and extensions to oxDNA presented here open up a variety of new potential applications for the model. With the introduction of an explicit salt-dependent term in the potential, the model can be used to simulate systems under physiological conditions, and to investigate the salt-dependent behaviour of DNA.  Also, the parametrization presented here allows a quantitative comparison to experiments run in a wide range of salt concentrations rather than just in the high-salt limit. The introduction of major-minor grooving adds detail to the model, and, combined with other small modifications, allows the use of oxDNA2 to accurately characterise the structural properties of large DNA nanostructures, such as DNA origami. In particular the helical pitch and the twist angles at nicks and junctions were fine-tuned to obtain a correct global twist for a test-case 3D origami structure. In the absence of definitive experimental values for these individual parameters we chose a combination that we deemed physically reasonable and that produces equilibrium structures that compare well with experimental ones. Finally, the sequence dependence in the model has been extended by introducing different interaction strengths for the AA and TT stacking. This change will be particularly useful for studying the effects of stacking in single strands or single-stranded sections (e.g.~hairpin loops) as poly-A and poly-T sections are often used as paradigmatic examples of strongly and weakly stacking sequences, respectively, and for modelling DNA nanostructures where poly-T loops are often used as flexible linkers.

There are some limitations associated with the new features of the model. The Debye-H\"uckel treatment of the electrostatics is perhaps an oversimplification, and we should not expect it to capture all of the complexities associated with the electrostatics for DNA, but it is not straightforward to think of a different approach that would be consistent with our level of coarse graining. In particular, the fact that hairpin stability is reduced faster than in experiment suggests that single strands may experience too much repulsion in oxDNA2 at low salt, and care should be taken in making predictions based on our models at low salt concentration if the system under investigation depends crucially on the thermodynamics of long single-stranded sections.

We are currently exploiting the improvements in the model's structural prediction to carry out investigations on DNA origami structures as well as on structures composed of multi-arm tiles\cite{maoTileReview2009}. At the same time, we are studying the salt dependent thermodynamics of a diverse set of systems, which would not have been possible without the introduction of salt dependence into the potential. The introduction of different strengths for the AA and TT stacking in the model allows us to better capture these effects, and in addition gives us more accurate sequence-dependent hairpin thermodynamics and kinetics, which are currently being exploited to further study hairpins.  We note that a simulation code implementing both the original oxDNA model and the new oxDNA2 model, including an implementation using GPUs,\cite{oxDNA_edge} is available for download from \url{dna.physics.ox.ac.uk}.

\section{Acknowledgements}

The authors are grateful to the Engineering and Physical Sciences Research Council for financial support. M. M. was supported by the Swiss National Science Foundation (Grant No.PBEZP2-145981). The authors thank Oxford's Advanced Research Computing, E-infrastructure South and the PolyHub virtual organization for computer time.

\section*{References}
\bibliography{citations}

\clearpage

\setcounter{figure}{0}
\makeatletter 
\renewcommand{\thefigure}{A\@arabic\c@figure}
\makeatother

\setcounter{table}{0}
\makeatletter 
\renewcommand{\thetable}{A\@arabic\c@table}
\makeatother

\onecolumngrid
\appendix

\renewcommand*{\theHtable}{\arabic{section}, \arabic{table}} 
\renewcommand*{\theHfigure}{\arabic{section}, \arabic{figure}} 

\section{Modifications to the oxDNA potential in oxDNA2}
\label{appendix:oxdna2 potential}
\subsection{Introducing an electrostatic term into the model potential}
\label{appendix:debye huckel}

We introduce an explicit electrostatic term into the model using Debye-H\"uckel theory. To aid computational efficiency when simulating the model, we ensure that the interaction goes to zero within a finite distance $r_{\rm cut, \rm DH}$, and we introduce a quadratic smoothing function so that the interaction goes to zero smoothly. The form of the electrostatic term is then
\begin{equation}
V_{\rm electrostatic}(r^{\text{b-b}},T,I) = \left\{
  \begin{array}{l l}
    V_{\rm DH}(r^{\text{b-b}},T,I) & \text{ if } r_{\rm smooth, \rm DH} > r^{\text{b-b}} \text{,}\\
    V_{\rm smooth}(r^{\text{b-b}},T,I) & \text{ if } r_{\rm cut, \rm DH} > r^{\text{b-b}} \ge r_{\rm smooth, \rm DH} \text{,} \\
    0 & \text{ otherwise,}
  \end{array}
\right.
\end{equation}
where $T$ is the temperature, $I$ is the salt concentration, and $r_{\rm smooth, \rm DH}$, the distance above which smoothing is introduced, is chosen to be equal to $3\lambda_{\rm DH}$ ($\lambda_{\rm DH}$ is defined below). The Debye-H\"uckel-like term, $V_{\rm DH}(r^{\text{b-b}},T,I)$, is given by
\begin{equation}
\label{appendix vdh}
V_{\rm DH}(r^{\text{b-b}},T,I) = \dfrac{(q_{\rm eff} e)^2}{4\pi \epsilon_0 \epsilon_{\rm r}}
\dfrac{
\exp{\left\{-r^{\text{b-b}}/\lambda_{\rm DH}(T,I)\right\}}
}{r^{\text{b-b}}} \text{,}
\end{equation}
where $q_{\rm eff}$ is the effective charge (which is equal to 1 in Debye-H\"uckel theory but which we choose to be 0.815 after fitting to experimental data), $e$ is the charge of an electron, $\epsilon_0$ is the permittivity of a vacuum, $\epsilon_r$ is the relative permittivity of water (which we choose to be 80), and $r^{\text{b-b}}$ is the distance between the pair of interacting backbone sites. $\lambda_{\rm DH}$ is the Debye screening length, and is given by
\begin{equation}
\label{appendix lambdadh}
\lambda_{\rm DH}(T,I) = \sqrt{\dfrac{\epsilon_0 \epsilon_{\rm r} k_{\rm B}T}{2N_{\rm A} e^2 I}}\:,
\end{equation}
where $k_{\rm B}$ is Boltzmann's constant and $N_{\rm A}$ is Avogadro's number. The quadratic smoothing term, $V_{\rm smooth}(r^{\text{b-b}},T,I)$, is given by
\begin{equation}
\label{dh smoothing}
V_{\rm smooth}(r^{\text{b-b}},T,I) = B(r^{\text{b-b}} - r_{\rm cut, \rm DH})^2 \text{,}
\end{equation}
and $B$ and $r_{\rm cut, \rm DH}$ are constants chosen such that $V_{\rm electrostatic}$ is smooth and differentiable. Note that the above definition implies that the cutoff radius for the Debye-H\"uckel term, $r_{\rm cut, \rm DH}$, depends on the salt concentration $I$ and the temperature $T$.

\subsection{Modifying the average pitch of the model duplex}
\label{appendix:pitch modification}
The model duplex pitch can be controlled by modifying the position of the minimum in the FENE bonded backbone potential, $\delta r^0_{\rm backbone}$, which has a small effect on the thermodynamics (which was refitted using the histogram reweighting method described in Section II C 2 of Ref.~\onlinecite{necitujoxRNA}) and little effect on the other properties of the model DNA. For oxDNA2, $\delta r^0_{\rm backbone}$ was set to give a global twist of zero for the N-type helix bundle due to Dietz \textit{et al.};\cite{3dDietz2009} the new and old values for $\delta r^0_{\rm backbone}$ are given in Table~\ref{table:oxDNA2 params}.

\subsection{Modifying the average coaxial stacking term in the oxDNA potential}
\label{appendix:coax change}
\begin{figure}
\begin{tikzpicture}
\begin{scope}
\draw node[anchor=north west] (schematic) {  \includegraphics[width=0.25\linewidth]{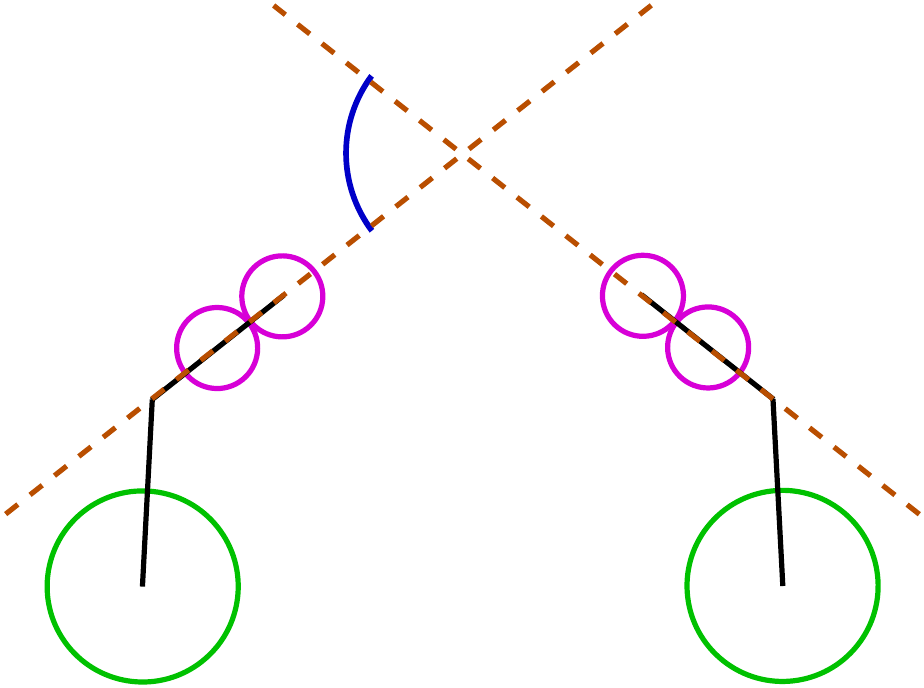}};
\draw (1.25cm,-0.65cm) node[anchor=north west] (theta1) {$\theta_1$};
\end{scope}
\end{tikzpicture}
  \caption{Schematic showing the angle $\theta_1$ between two nucleotides. Each nucleotide is represented by three circles joined by a line; the large solid circles represent the backbone sites, while the small solid circles represent the stacking (closer to the backbone) and hydrogen-bonding (at the end of the nucleotide) sites. In this paper we modify the term dependent on $\theta_1$ for the coaxial stacking interaction in the oxDNA model, in order to reduce the overtwist across nicks and junctions.}
  \label{fig:coax theta1}
\end{figure}

We modified the coaxial stacking term in the oxDNA potential in order to change the twist angle across a nick and across an origami junction, while changing the thermodynamic properties of a nicked duplex and other motifs as little as possible. This was achieved by moving the position of the minimum with respect to the angle $\theta_1$ (shown in Fig.~\ref{fig:coax theta1}) of the coaxial stacking term, $\theta_{\rm coax, 1}^0$, and by increasing the well depth, $k_{\rm coax}$, to make the minimum more energetically favourable. Increasing the well depth was necessary because it is less favourable for a pair of coaxially stacking nucleotides in a nicked duplex to stack with this tighter angle; a stronger overall coaxial stacking interaction compensates for this effect. The changes to the model are shown in Table~\ref{table:oxDNA2 params}.

\begin{table*}
  \begin{tabular}{c c c}
    \hline
    \hline
    Parameter & oxDNA & oxDNA2 \\ \hline
    $\gamma$/degrees & $0$ & $20$ \\
    $q_{\rm eff}$ & $0$ & $0.815$ \\
    $\epsilon_{\rm r}$ & - & 80 \\
    $k_{\rm coax}$ & $46$ & $58.5$ \\ 
    $\theta_{\rm coax, 1}^0$ & $\pi -0.60$ & $\pi -0.25$ \\
    $A_{\rm coax,1}$ & - & $40$ \\
    $B_{\rm coax,1}$ & - & $\pi - 0.025$ \\
    $\epsilon_{\rm HB}$ & $1.077$ & $1.0678$ \\ 
    $\epsilon_{\rm stack}$ & $1.3448+2.6568k_{\rm B}T$ & $1.3523+2.6717k_{\rm B}T$ \\
    $\delta r^0_{\rm backbone}$ & $0.7525$ & $0.7564$ \\ 
    \hline
    \hline
  \end{tabular}
  \caption{Comparison of model parameters between oxDNA and oxDNA2. All other parameters are unchanged in oxDNA2. All angles except $\gamma$ are given in radians, all lengths are defined with respect to a reduced length scale (1 unit = 8.518$\rm \r A$) and all energies are defined with respect to a reduced temperature ($k_{\rm B}T$ = 0.1 corresponding to 300K). $\gamma$ is the angle between the line from the helix axis to the backbone site and the line from the helix axis to the stacking site (see Fig.~\ref{fig:mmgroove}), which defines the backbone site position in oxDNA2. $q_\text{eff}$ is the effective charge used for the Debye-H\"uckel treatment introduced in the new model for the electrostatic interactions due to the negatively-charged phosphate groups in the DNA backbone. $\epsilon_{\rm r}$ is the relative permittivity of water which is used for the Debye-H\"uckel treatment. $k_{\rm coax}$ determines the overall strength of the coaxial stacking interaction while $\theta^0_{\rm coax, 1}$ specifies the position of the minimum with respect to the angle $\theta_1$ of the coaxial stacking potential. The values for $A_{\rm coax, 1}$ and $B_{\rm coax, 1}$ are chosen to ensure that the coaxial stacking potential is smooth and differentiable. $\epsilon_{\rm HB}$ and $\epsilon_{\rm stack}$ determine the well depths for the hydrogen bonding and stacking interactions respectively. $\delta r^0_{\rm backbone}$ gives the position of the minimum of the FENE bonded backbone potential.}

  \label{table:oxDNA2 params}
\end{table*}

The altered optimal angle $\theta_{\rm coax,1}^0$ requires an alteration of the modulation function involving $\theta_1$ for coaxial stacking. In the original oxDNA, the $\theta_1$ modulation is given by $f_4(\theta_1) + f_4(2 \pi -\theta_1)$, where $f_4$ is defined in Ref.~\onlinecite{ouldridgeThesis}. The second term was introduced to ensure that the potential is differentiable at $\theta_1=\pi$, where the gradients of the two terms cancel. In oxDNA, this additional term had a very small effect on the numerical value of the potential for $\theta_1 < \pi$. Using the same trick in oxDNA2, however, is problematic, because $\theta_{\rm coax,1}^0$ is much closer to $\pi$. The second term, $f_4(2 \pi -\theta_1)$, would then be substantial for $\theta_1 < \pi$ and would distort the modulating function in an undesired fashion. To avoid this, we replace the second term, $f_4(2 \pi -\theta_1)$, with a new function $f_6(\theta_1)$:

\begin{equation}
f_6\left(\theta \right) =
\begin{cases}
  \hfill \frac{A}{2} (\theta - B)^2 \text{.}                     \hfill & \text{ if $\theta \geq B$,} \\
  \hfill 0                                                       \hfill & \text{ otherwise.}
\end{cases}
\end{equation}
The $f_4(\theta_1)+f_4(2\pi - \theta_1)$ factor in the coaxial stacking term is then replaced by $f_4(\theta_1)+f_6(\theta_1)$. The values for $A_{\rm coax, 1}$ and $B_{\rm coax, 1}$ are chosen to ensure that the coaxial stacking potential is smooth and differentiable; they are specified in Table~\ref{table:oxDNA2 params}.

For oxDNA2, we made an additional change to the coaxial stacking potential: we removed the $f_5(\cos (\phi_3), a_{\rm coax, 3'}, \cos (\phi_3)^*_{\rm coax})$ and $f_5(\cos (\phi_4), a_{\rm coax, 4'}, \cos (\phi_4)^*_{\rm coax})$ terms. These terms had allowed only right-handed blunt-ended stacking, disallowing left-handed blunt-ended stacking; however, since the development of the original oxDNA model, this constraint has been deemed unnecessary, as the blunt-ended coaxial stacking interaction was found to be likely to be too weak compared to experiment,\cite{CDMnematicDNA} and there is no experimental evidence indicating that left-handed blunt-ended coaxial stacking is not possible. 

Incorporating the changes described above, the final form for the oxDNA2 coaxial stacking term is

\begin{align}
V_{\rm coax\_stack} &= f_2(\delta r_{\rm stack}, k_{\rm coax}, \delta r^0_{\rm coax}, \delta^{c,\rm low}_{\rm coax}, \delta r^{c,\rm high}_{\rm coax}, \delta r^{\rm low}_{\rm coax}, \delta r^{\rm high}_{\rm coax}) f_4(\theta_4, a_{\rm coax, 4}, \theta^0_{\rm coax, 4}, \Delta \theta_{\rm coax, 4}^*) \nonumber \\
&\qquad {} \times (f_4(\theta_1, a_{\rm coax, 1}, \theta^0_{\rm coax, 1}, \Delta \theta^*_{\rm coax, 1}) + f_6(\theta_1, A_{\rm coax, 1}, B_{\rm coax, 1})) \nonumber \\
&\qquad {} \times (f_4(\theta_5, a_{\rm coax, 5}, \theta^0_{\rm coax, 5}, \Delta \theta^*_{\rm coax, 5}) + f_4(\pi - \theta_5, a_{\rm coax, 5}, \theta^0_{\rm coax, 5}, \Delta \theta^*_{\rm coax, 5})) \nonumber \\
&\qquad {} \times (f_4(\theta_6, a_{\rm coax, 6}, \theta^0_{\rm coax, 6}, \Delta \theta^*_{\rm coax, 6}) + f_4(\pi - \theta_6, a_{\rm coax, 6}, \theta^0_{\rm coax, 6}, \Delta \theta^*_{\rm coax, 6})) \text{.}
\end{align}

\begin{figure}
  \begin{tikzpicture}
    \begin{scope}
      \draw node[anchor=north west] (schematic) {\includegraphics[width=0.4\linewidth]{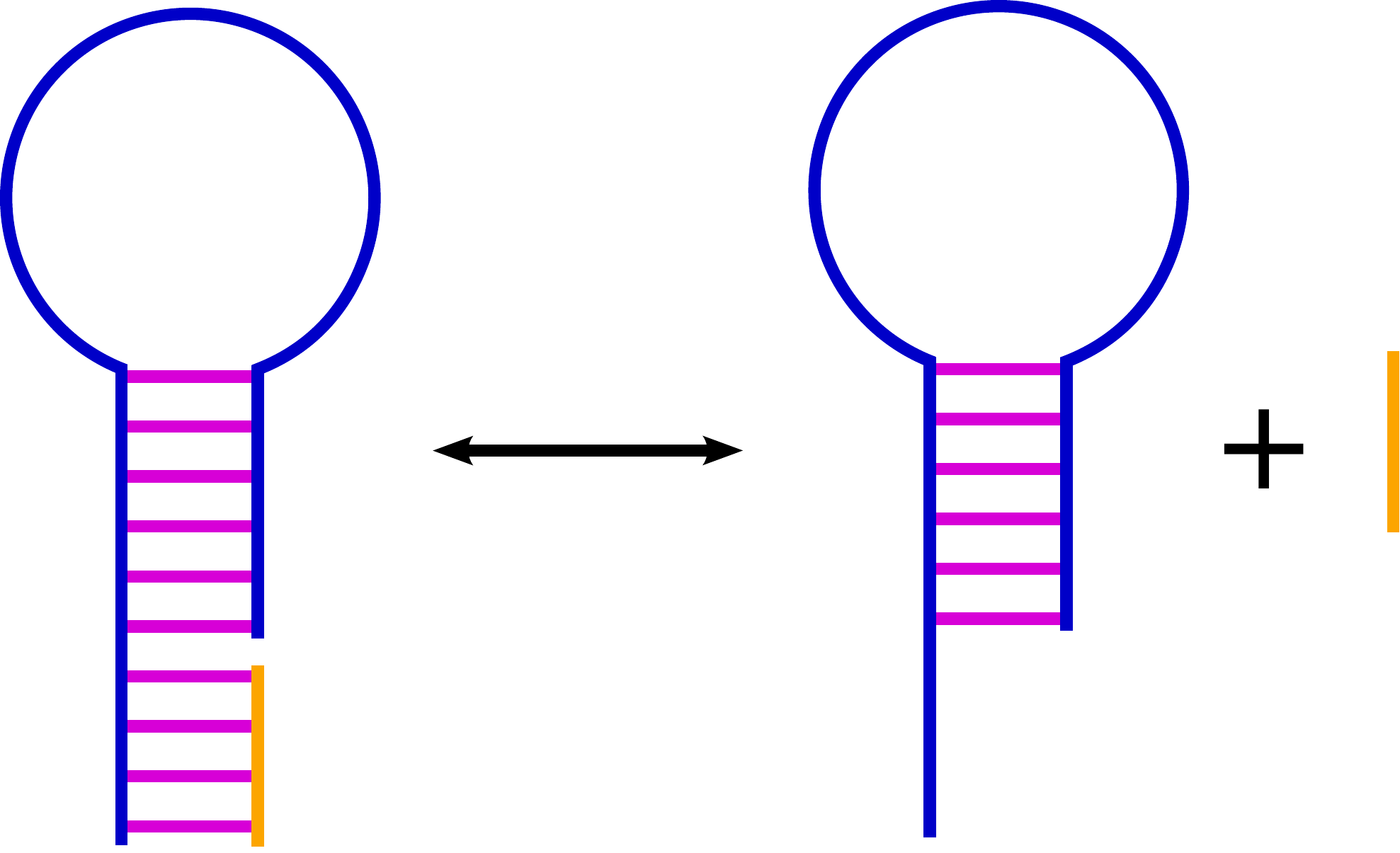}};
    \end{scope}
  \end{tikzpicture}
  \caption{A schematic showing the transition important for the calculation of $\Delta T_{\rm m, \rm hairpin\_stem,X\text{mer}}$ (see Table \ref{table:coax thermo}), which is given by $\Delta T_{\rm m, \rm hairpin\_stem,X\text{mer}} = T_{\rm m, \rm hairpin\_stem,X\text{mer}} - T_{\rm m, X\text{mer}}$, where $T_{\rm m, \rm hairpin\_stem,X\text{mer}}$ is the melting temperature of a X-bp duplex adjacent to a hairpin (the schematic shows this system for the X\,$=4$ case) and $T_{\rm m, X\text{mer}}$ is the melting temperature of a X-bp duplex.}
  \label{fig:coax hairpin stab}
\end{figure}

Table \ref{table:coax thermo} gives a comparison of the thermodynamic properties of the motifs relevant to the coaxial stacking term in oxDNA and oxDNA2. The free-energy change upon stacking across a nick, $\Delta G_{\rm nick\_stack}$, is slightly more negative in oxDNA2, indicating that a nicked duplex stacks slightly more strongly across the nick in oxDNA2 than in the original oxDNA. In both cases the stacking is stronger than is seen in experiment ($-2.62k_{\rm B}T$ in Ref.~\onlinecite{yakovchuk2006} -- a discussion comparing the value of $\Delta G_{\rm nick\_stack}$ measured for the original oxDNA with experimental results is given in Ref.~\onlinecite{ouldridgeThesis}). The stabilising effect of a hairpin stem on duplex hybridisation (see Fig.~\ref{fig:coax hairpin stab}, $\Delta T_{\rm m, \rm hairpin\_stem,X\text{mer}}$, is found to be slightly stronger than for the original oxDNA, and is on average in better agreement with the stability predicted by the SantaLucia model. Finally, dimerisation of two duplexes through blunt-ended coaxial stacking is found to be less favourable in oxDNA2 than in oxDNA. This is despite the mitigating effect of removing the $f_5$ terms in the coaxial stacking interaction, which strengthens blunt-ended stacking. A previous study\cite{CDMnematicDNA} has shown that oxDNA already underestimated the stability of blunt-ended coaxial stacking.

\begin{table*}
  \begin{tabular}{c c c c}
    \hline
    \hline
    Measurement & oxDNA & oxDNA2 & SantaLucia \\ \hline
    $\Delta G_{\rm nick\_stack}$ & $-4.3k_{\rm B}T$ & $-4.4k_{\rm B}T$ & - \\ 
    $\Delta T_{\rm m, \rm hairpin\_stem,6\text{mer}}$ & $+9.9\rm K$ & $+10.6\rm K$ & $+11.7\rm K$ \\
    $\Delta T_{\rm m, \rm hairpin\_stem,7\text{mer}}$ & $+7.7\rm K$ & $+8.3\rm K$ & $+8.5\rm K$ \\
    $\Delta T_{\rm m, \rm hairpin\_stem,8\text{mer}}$ & $+6.3\rm K$ & $+6.9\rm K$ & $+6.4\rm K$ \\
    $\Delta G_{\rm blunt\_ended\_stacking}$ & $+4.9k_{\rm B}T$ & $+5.3k_{\rm B}T$ & - \\
    \hline
    \hline
  \end{tabular}
  \caption{Comparison of some thermodynamic quantities that depend on the coaxial stacking interaction in oxDNA and oxDNA2. $\Delta G_{\rm nick\_stack} = \Delta G_{\rm stacked} - \Delta G_{\rm unstacked}$ is the free-energy change upon stacking across a nick for a 20-bp duplex with a nick at the centre at $37^\circ$\,C. $\Delta T_{\rm m, \rm hairpin\_stem,6\text{mer}} = T_{\rm m, \rm hairpin\_stem,6\text{mer}} - T_{\rm m, 6\text{mer}}$ gives the difference in melting temperature at a concentration of $3.3\times 10^{-4}$\,M between a 6-bp duplex adjacent to a hairpin and a 6-bp duplex (see Fig.~\ref{fig:coax hairpin stab}), and $\Delta T_{\rm m, \rm hairpin\_stem,7\text{mer}}$ and $\Delta T_{\rm m, \rm hairpin\_stem,8\text{mer}}$ have analogous definitions. $\Delta G_{\rm blunt\_ended\_stacking}$ is the free-energy change upon dimerisation of two 6-bp duplexes through blunt-ended coaxial stacking at a monomer concentration of $5.37$\,mM and at $19.85^\circ$\,C. $\Delta G_{\rm nick\_stack}$ was computed from a simulation of $1.9\times 10^{11}$ VMMC steps, each $\Delta T_{\rm m, \rm hairpin\_stem,X\text{mer}}$ from simulations of $8\times 10^{10}$ VMMC steps, and $\Delta G_{\rm blunt\_ended\_stacking}$ from a simulation of $2.4\times 10^{11}$ VMMC steps.}
  \label{table:coax thermo}
\end{table*}

\subsection{Modifying the hydrogen bonding and stacking interaction strengths}
\label{appendix:thermodynamic reparameterisation}
Model duplexes were found to fray slightly less often when different widths for the major and minor duplex grooves were introduced compared to the equal-groove-width case, which led to a slight raising of the duplex melting temperatures. In addition, the change to the position of the minimum in the FENE bonded backbone potential, $\delta r^0_{\rm backbone}$, was found to destabilise the duplex state, slightly lowering the duplex melting temperatures. In the process of the reparameterisation of oxDNA to incorporate these two changes, the duplex melting temperatures were reset on the introduction of each change by making small modifications to the stacking and hydrogen bonding strengths using the histogram reweighting method described in Section II C 2 of Ref.~\onlinecite{necitujoxRNA}, giving an agreement with experimental melting temperatures that was just as good as before. The new parameters are specified, and compared to the old parameters, in Table~\ref{table:oxDNA2 params}, while Fig.~\ref{fig:duplex thermo} gives a comparison of the duplex thermodynamics with the old and new versions of the model.

\section{Simulation methods}
\subsection{Measuring the global twist of an origami}
\label{appendix:global twist}
We measure the global twist of the origami due to Dietz \textit{et al.}\cite{3dDietz2009} in the following way: The origami (for example, the L-type, although this applies equally well to all three types) can be thought of as a set of antiparallel double helices on a hexagonal lattice joined together by crossovers. The overall shape of this structure is roughly that of a (possibly twisted) rectangular prism. To measure the twist, we consider the two faces of the rectangular prism which are opposite each other and approximately normal to the axes of the double helices. For each face, we define a vector along its long edge, and we superimpose these two vectors into a plane perpendicular to the average of the helix axes. We then define the global twist as the angle between these two vectors.

To avoid end effects, the top and bottom rows of helices (3 helices in the top row and 3 helices in the bottom row) were excluded for this analysis, as were approximately 22 base pairs from each end of every double helix.

\subsection{Measuring the torsional stiffness of a duplex in the oxDNA model}
\label{appendix:torsional stiffness}

The torsional stiffness of a duplex in oxDNA2 was measured according to the scheme outlined by Matek \textit{et al.}\cite{matekPlectoneme} In summary, this method involves pulling a duplex while also twisting it, and measuring the torque required to impose a given twist. In this case, a 60-bp duplex was pulled at a constant force of 30\,pN and held with virtual traps which imposed a helical pitch different from the equilibrium pitch. For each salt concentration, simulations were run with imposed pitches of 9.8, 10.1, 10.4, 10.7, 11.0, and 11.3 bp/turn, with each simulation run for at least $2.6 \times 10^9$ MD steps. The torque observed was plotted as a function of imposed twist (this is called a torque response curve), and the gradient of the linear region used to approximate the effective torsional stiffness.

The torsional stiffness $C_{\rm eff}$ is computed as
\begin{equation}
C_{\rm eff} = \frac{\Delta \Gamma}{\Delta \sigma}\frac{a_0}{\theta_0} \text{,}
\end{equation}
where $\Delta \Gamma$ is the change in torque, and $\Delta \sigma$ the change in superhelical twist density, measured in the linear regime of the torque response curve, $a_0$ is the double helical rise for a relaxed duplex and $\theta_0$ is the twist angle across a base-pair step for a relaxed duplex, measured in radians. According to Moroz and Nelson,\cite{morozTwisting1997} the effective torsional stiffness $C_{\rm eff}$ measured by twisting a duplex under tension can be related to the true torsional stiffness $C_0$ by
\begin{equation}
C_{\rm eff} = C_0 \left\{ 1  - \frac{C_0}{4B_0} \sqrt{\frac{k_{\rm B}T}{B_0 F}} \right\} \text{,}
\end{equation}
where $F$ is the linear force and $B_0$ is the bending stiffness. Our simulations are in the high-force regime, where $C_{\rm eff}$ should be a good approximation for $C_0$.

\subsection{Details of the simulation protocols}
\label{appendix:simulation details}

The simulation results presented in this paper were obtained using either a virtual move Monte Carlo (VMMC) algorithm or a molecular dynamics (MD) algorithm with a Brownian thermostat. Note that, where VMMC steps are reported in the main text, they refer to the number of \textit{accepted} moves during the simulation. Details of the simulation parameters used are given in Table~\ref{table:simulation parameters}.

For the duplex free-energy profiles, yields and melting temperatures (Fig.~\ref{fig:duplex thermo} and Fig.~\ref{fig:duplex melting}) we used umbrella sampling with an order parameter defined in terms of the number of native base pairs (0 to $n$ where $n$ is the number of base pairs in a fully formed duplex). A base pair was defined as being formed if the potential's hydrogen-bonding energy term for that pair was lower than $-0.596$\,$\rm kcal mol^{-1}$. For the hairpin melting temperature calculations (Fig.~\ref{fig:hairpins}, Fig.~\ref{fig:low salt hairpins} and Fig.~\ref{fig:hairpins df performance}) we used an analogous order parameter for the native base pairs in the hairpin stem.

The oxDNA2 results reported in Fig.~\ref{fig:hairpins df performance} were computed using thermodynamic integration as described in Section~\ref{section:aa/tt}. For each oxDNA2 data point shown in the figure, first an umbrella sampling simulation of each hairpin (sequences shown in Table~\ref{table:experiment hairpin sequence}) was carried out for a version of the model with $x=0$, where $x$ is given by Eq.~\ref{eq:x def}, and $q_{\rm eff}=0$ in order to find the reference free energy ($F^{(0)}_{\alpha}$ in Eq.\,\ref{eq:TI2}). The simulations were run for $6\times 10^{11}$ VMMC steps and used the parameters specified in Table~\ref{table:simulation parameters}. The integrals in Eq.\,\ref{eq:TI2} were discretised into 15 gridpoints for the integral in $q^{\prime}_{\rm eff}$ and 10 gridpoints for the integral in $x$. Then simulations to compute these integrals were run for roughly $6\times 10^7$ VMMC steps for each grid point, so that each data point in the figure required 25 grid point simulations, with one umbrella sampling simulation per hairpin.

\begin{table*}
  \begin{tabular}{c c c c c c c c c c c c c c}
    \hline
    \hline
    & \multicolumn{9}{c}{Simulation description} \\
    \cline{2-10}
    Parameter & Standard & Bulge & Nick stack. & Hairpin stab. & Blunt coax & Hairpin $\delta F$ & $C_{\rm torsional}$ & $l_{\rm p}$ & Helix bundle twist \\ \hline
    Algorithm & VMMC & VMMC & VMMC & VMMC & VMMC & VMMC &MD & MD & MD \\ 
    $\delta_{\rm trans}$ & $0.11$ & $0.1$ & $0.03$ & $0.15$ & $0.22$ & $0.11$ & - & - & - \\
    $\delta_{\rm rot}$ & $0.22$ & $0.2$ & $0.15$ & $0.15$ & $0.22$ & $0.22$ & - & - & - \\
    max. cluster size & - & - & - & - & $12$ & - & - & - & - \\
    Verlet skin & $1$ & $1$ & $1$ & $1$ & $1$ & $1$ & $0.05$ & $0.1$ & $0.05$ \\
    $\delta t_{\rm step}$ & - & - & - & - & - & - & $0.005$ & $0.005$ & $0.005$ \\
    diffusion coefficient & - & - & - & - & - & - & $2.5$ & $2.5$ & $2.5$ \\
    $T$ & $T_{\rm m}$ & $T_{\rm m}$ & $310.15\rm K$ & $T_{\rm m}$ & $293\rm K$ & $295.6\rm K$ & $296.15\rm K$ & $296.15\rm K$ & $296.15\rm K$ \\
    \hline
    \hline
  \end{tabular}
  \caption{The results in this paper were computed using simulations of oxDNA2 with these parameters. The ``standard'' parameters were used for the following calculations: $\Delta T_{\rm m}$ for a terminal mismatch and $\Delta T_{\rm m}$ for an internal mismatch (Table~\ref{table:other motifs}); all hairpin $T_{\rm m}$ calculations (Fig.~\ref{fig:hairpins} and Fig.~\ref{fig:low salt hairpins}); and all duplex $T_{\rm m}$ calculations (Fig.~\ref{fig:duplex thermo}). The ``bulge'' parameters were used for the $\Delta T_{\rm m}$ calculations for a bulge motif (Table~\ref{table:other motifs}), the ``nicked stacking'' parameters were used for the $\Delta G_{\rm nick\_stack}$ calculation (Table~\ref{table:coax thermo}), the ``hairpin stem stab.'' parameters were used for the $\Delta T_{\rm m, hairpin\_stem,Xmer}$ calculations (Table~\ref{table:coax thermo}), the ``blunt coax'' parameters were used for the calculation of $\Delta G_{\rm blunt\_ended\_stacking}$ (Table~\ref{table:coax thermo}), the ``hairpin $\delta F$'' parameters were used for the initial free energy calculation for the thermodynamic integration to find the relative stability of the hairpins  shown in Fig.~\ref{fig:hairpins df performance}, and to find the radius of gyration ($R_{\rm g}$) of ssDNA (Fig.~\ref{fig:rg}), the ``$C_{\rm torsional}$'' parameters were used for the calculation of the torsional stiffness (Fig.~\ref{fig:mechanical properties}(b)), the ``$l_{\rm p}$'' parameters were used for the calculation of the persistence length (Fig.~\ref{fig:mechanical properties}(a)) and the pitch and rise (Fig.~\ref{fig:pitch and rise}), and the ``helix bundle twist'' parameters were used for the calculation of the global twist of the helix bundle origami structures (Table~\ref{table:monolith twist table}). $\delta_{\rm trans}$ and $\delta_{\rm rot}$ are the standard deviation of the normally distributed move size for the translational and rotational VMMC moves respectively, max. cluster size is the size of the biggest cluster that VMMC is permitted to choose for a trial move, Verlet skin is the thickness of the Verlet skin used for Verlet lists, $\delta t_{\rm step}$ is the size of the simulation time step in MD, diffusion coefficient gives the diffusion coefficient used for a single nucleotide, and $T$ is the temperature of the heat bath ($T_{\rm m}$ indicates that simulations were run at or close to the melting temperature for that system). All numbers are given in oxDNA units (1 oxDNA length unit = 0.8518\,nm, 1 oxDNA time unit = 3.03\,ps), unless given explicitly (angles are quoted in radians).}
  \label{table:simulation parameters}
\end{table*}

\section{Experimental Methods}
\label{appendix:experimental hairpin}
\subsection{DNA Hairpin Design}
The hairpin constructs consist of a loop with different lengths (21 or 31 nucleotides) and sequences (adenosine or thymine), and six base pairs in the stem, as shown in Table~\ref{table:experiment hairpin sequence}. 

\begin{table*}
  \begin{tabular}{c c}
    \hline
    \hline
    Name & Hairpin Sequence \\ \hline
    $\rm T_{21}\text{-}W$ & $\rm 5^{\prime}\text{-}TGGGTT\text{-}(T)_{21}\text{-}AACCCA\text{-}3^{\prime}$ \\ 
    $\rm T_{31}\text{-}VW$ & $\rm 5^{\prime}\text{-}TGGATT\text{-}(T)_{31}\text{-}AATCCA\text{-}3^{\prime}$ \\ 
    $\rm A_{21}\text{-}W$ & $\rm 5^{\prime}\text{-}TGGGTT\text{-}(A)_{21}\text{-}AACCCA\text{-}3^{\prime}$ \\ 
    $\rm A_{31}\text{-}VW$ & $\rm 5^{\prime}\text{-}TGGATT\text{-}(A)_{31}\text{-}AATCCA\text{-}3^{\prime}$ \\ 
    \hline
    \hline
  \end{tabular}
  \caption{DNA hairpins used for oxDNA validation. Note that the experimental system included an additional dangling duplex attached to the hairpin (denoted ``dsDNA part'' in Fig.~\ref{fig:experiment hairpin schematic}). For the oxDNA2 simulations, we used a truncated (13-bp) duplex in order to capture any effects due to the duplex interfering with the hairpin while avoiding the larger computational cost of simulating the full duplex.}
  \label{table:experiment hairpin sequence}
\end{table*}

\begin{figure}
  \includegraphics[width=0.8\linewidth]{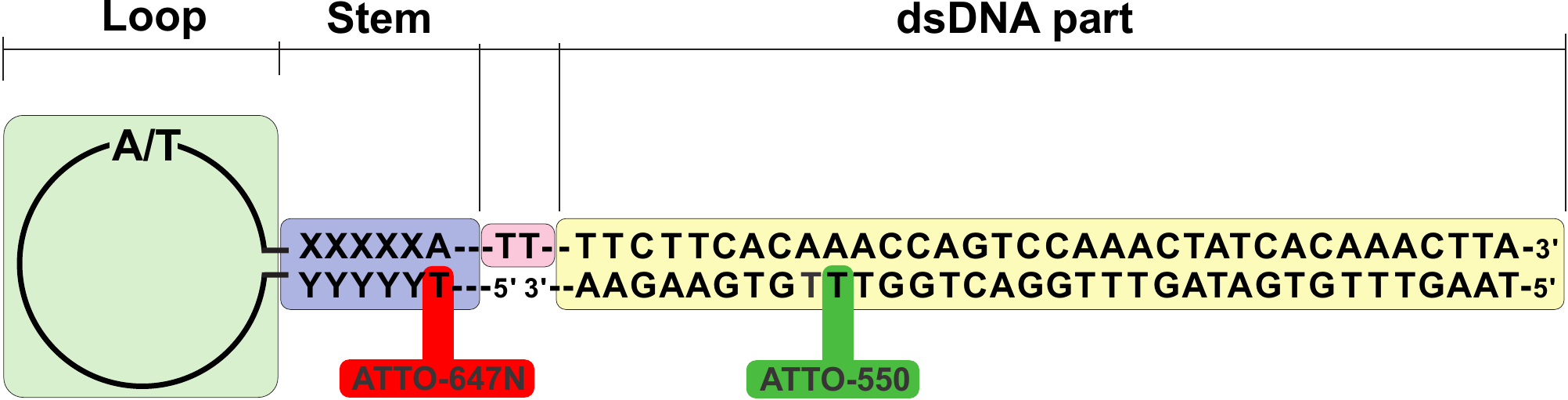}
  \caption{Design of the DNA hairpin constructs used to parameterise the oxDNA model. The hairpin sequences and labeling positions were designed to minimize possible interfering interactions between the fluorophores when the hairpins were closed, and the TT spacer was introduced to minimize interactions between the stem and the duplex.}
  \label{fig:experiment hairpin schematic}
\end{figure}

A 35-bp long, double-stranded DNA was used to keep the hairpin section away from the coverslip and the origami surfaces. The hairpins were labeled with donor and acceptor fluorophores (ATTO-550 and ATTO-647N, respectively), positioned such that the hairpin open state yielded low FRET values ($E$) and the closed state yielded high $E$ values (Figure~\ref{fig:experiment hairpin schematic}).

\subsection{Annealing Procedures and Experimental Setup}
The top and bottom strands were annealed at $94^{\circ}$C (1.5\,$\upmu$M in 10\,$\upmu$L TE-NaCl buffer [10\,mM Tris, pH 8.0, 1\,mM EDTA], and 100\,mM NaCl) and then gradually cooled (30 min) to room temperature (using PCR).

The measurements were performed on a diffusion-based single-molecule FRET-ALEX setup, as described elsewhere.\cite{EyalConformationalDynamicsOfHairpins} In brief, a green CW laser beam (532\,nm, CL532-025-L, Crystal Laser, Reno, NV, USA) was aligned/misaligned into a single-mode fiber using an acousto-optic modulator (AOM; R23080-2-LTD, Neos Technologies, Melbourne, FL, USA), alternating with a red diode laser (640\,nm, 1069417, Coherent, Auburn, CA, USA) that was electronically switched on/off. The AOM and the red laser were computer controlled with a 12.4-$\upmu$s on-time, a 12.6-$\upmu$s off-time, and a phase shift of 12.5\,$\upmu$s. The intensity rise and fall times were less than 50\,ns, and there was no time overlap between the lasers. The laser beams were combined by a dichroic mirror (Z532RDC, Chroma, Bellows Falls, VT, USA) and coupled into a single-mode fiber (P1-460A-FC-2, Thorlabs, Newton, NJ, USA). The laser intensities were tuned such that the doubly labeled species would yield a stoichiometry, $S$, of roughly 0.5 (90\,$\upmu$W for the green laser and 35\,$\upmu$W for the red laser, measured after the fiber while alternating). After collimation (objective PLCN10×/0.25, Olympus America, Melville, NY, USA), the combined green and red beams were introduced into a commercial inverted microscope (IX71, Olympus America) and focused about 70\,$\upmu$m inside the sample solution with a water-immersion objective (NA 1.2, 60$\times$, Olympus America). The emitted fluorescence was separated from the excitation light by a dichroic mirror (ZT532/638RPC, Chroma), focused into a 100-$\upmu$m pinhole (P100S, Thorlabs), re-collimated, split by a second dichroic mirror (FF650-Di01, Semrock, Lake Forest, IL, USA), passed through filters (band-pass filter, Semrock FF01-580/60, for the donor channel and a long-pass filter, Semrock BLP01-635R, for the acceptor channel), and focused into two single-photon avalanche photodiodes (SPAD; SPCM-AQRH-13, Perkin-Elmer Optoelectronics, Fremont, CA, USA). The TTL signals of the two SPADs were recorded as a function of time by a 12.5-ns resolution counting board (PCI-6602, National Instruments, Austin, TX, USA) and in-house prepared LabView acquisition software.

\subsection{Measurement Solution and Conditions}
Measurements were carried out on diluted samples (40\,$\upmu$L of 1\,pM hairpins) that were placed on a coverslip and sealed with silicone and an upper coverslip. The bottom coverslip was KOH-treated to prevent sticking to the surface, sonicated for 15 min in 1\,M KOH solution, thoroughly washed with distilled water, and dried in air. The measurement buffer comprised 10\,mM Tris pH 8.0, 1\,mM EDTA, 10\,$\upmu$g/mL bovine serum albumin (BSA, Sigma-Aldrich) to reduce sample sticking, 2\,mM Trolox (Sigma-Aldrich) to reduce photo-bleaching and photo-blinking of the dyes, and indicated NaCl concentrations. Data were collected for over 30 minutes, and temperature was maintained at $22.5^{\circ}$\,C.

\subsection{Data analysis}
\begin{figure}
  \includegraphics[width=0.5\linewidth]{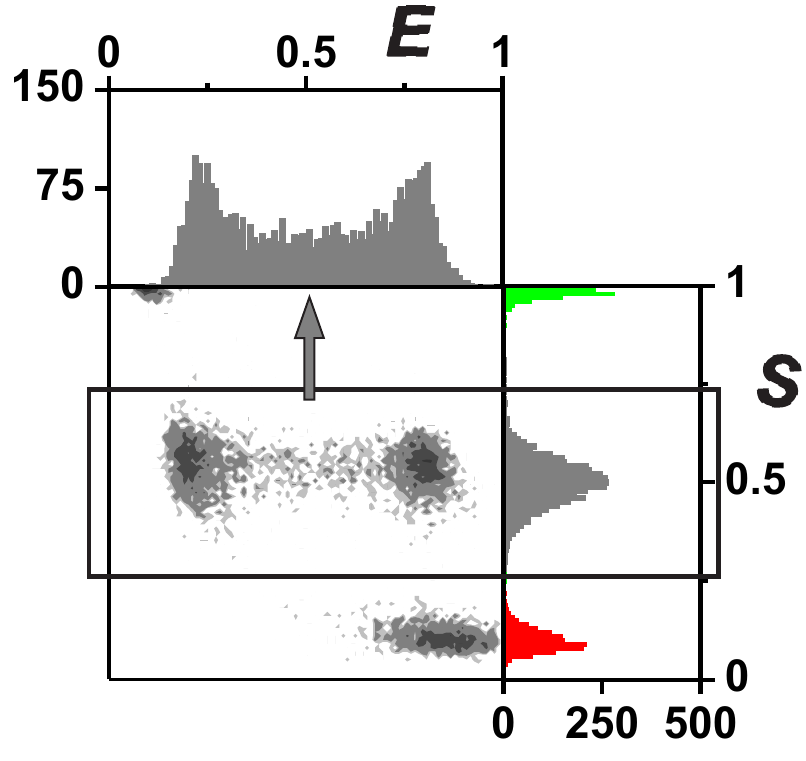}
  \caption{Typical data for hairpin two-dimensional $E/S$-histogram and $E$- and $S$- one-dimensional histograms.  The specific data refers to A31-VW measured in buffer containing 100 mM NaCl. Donor-only (green) and acceptor-only (red) populations are rejected by $S$-ratio. The populations with $S \approx 0.5$ (inside black rectangle) were used to obtain the final $E$-histogram.\cite{EyalHairpinDynamicsOrigami}}
  \label{fig:experiment hairpin histogram}
\end{figure}

Data analysis was performed with the in-house written LabView software as described elsewhere.\cite{EyalHairpinDynamicsOrigami} The beginnings and endings of bursts were determined by the all-photons-burst-search (APBS, parameters: $L$ = 200, $M$ = 10, and $T$ = 500 \,$\upmu$s. For each burst, $E$ and $S$ were calculated according, respectively, binned (0.01 bin size), and plotted on one-dimensional $E$ and $S$ histograms and on a two-dimensional $E/S$ histogram.\cite{EyalHairpinDynamicsOrigami} Donor-only and acceptor-only populations were rejected by ALEX. The Obtained $E$-histogram showed two distinct peaks and was assumed to have quasi two-state dynamics\cite{EyalHairpinDynamicsOrigami} (Figure~\ref{fig:experiment hairpin histogram}).

The opening and closing rates were extracted from the shape of the E-histogram using dynamic Probability Distribution Analysis.\cite{EyalConformationalDynamicsOfHairpins, WeissShotNoiseLimitedFRET}

\subsection{Experimental Data}
\begin{figure}
  \includegraphics[width=0.5\linewidth]{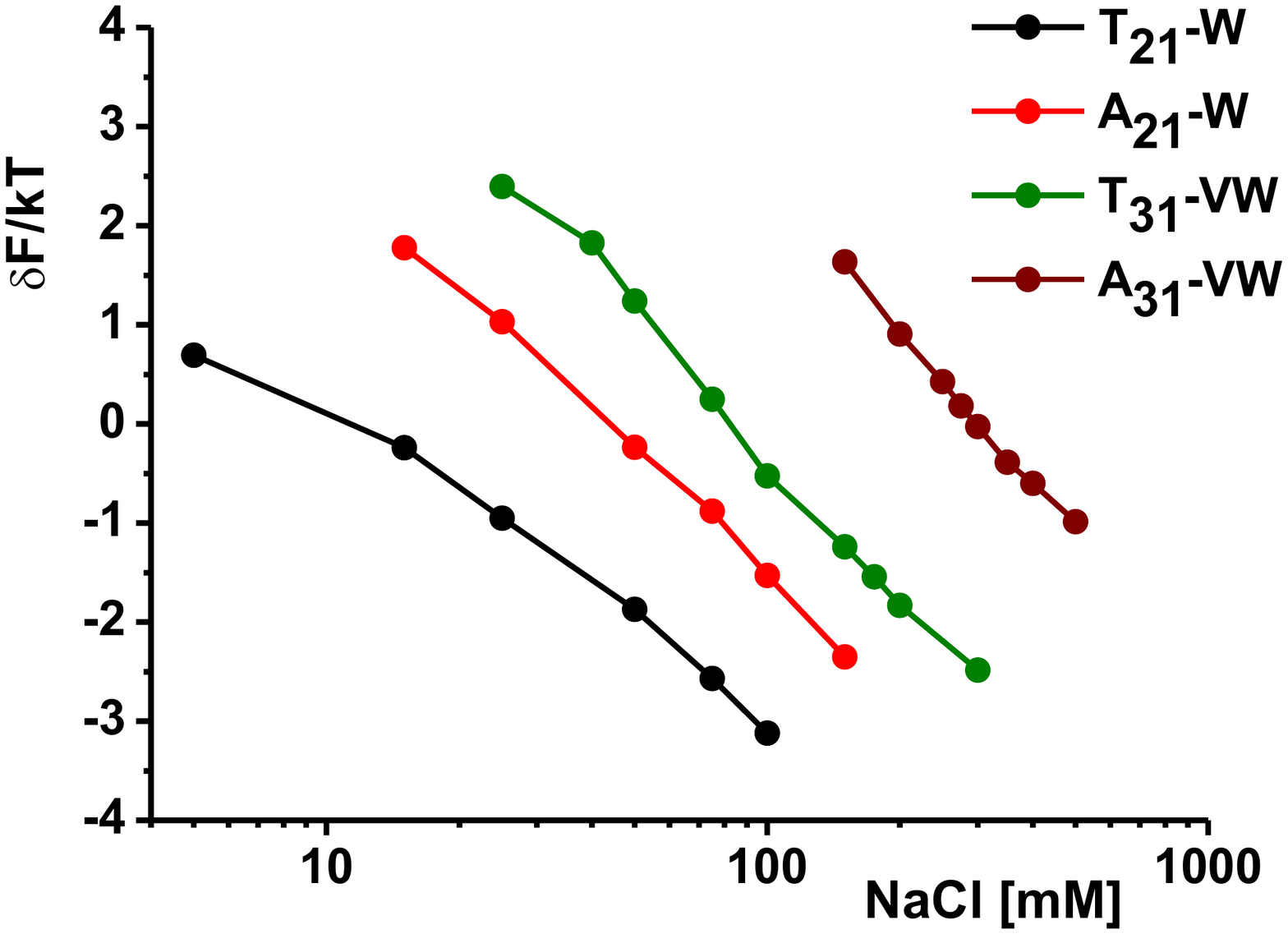}
  \caption{The free energy difference between the bound and unbound state ($\delta F$) experimentally determined for the four hairpins.}
  \label{fig:experimental hairpin deltaF}
\end{figure}

The results of the hairpin kinetics experiments are shown in Fig.~\ref{fig:experimental hairpin deltaF}.

\end{document}